\documentclass{article}
\usepackage[preprint]{neurips_2024}

\usepackage[utf8]{inputenc}
\usepackage[T1]{fontenc}
\usepackage{hyperref}
\usepackage{url}
\usepackage{booktabs}
\usepackage{amsmath,amssymb,amsthm,mathtools}
\usepackage{graphicx}
\usepackage{xcolor,colortbl}
\usepackage{multirow}
\usepackage{enumitem}
\usepackage{tikz}
\usetikzlibrary{positioning,arrows.meta,fit,backgrounds,decorations.pathreplacing,calc,shapes.geometric}
\usepackage{pgfplots}
\pgfplotsset{compat=1.18}
\usepgfplotslibrary{fillbetween}
\usepgfplotslibrary{statistics}
\usepgfplotslibrary{groupplots}
\usepackage{caption}
\usepackage{subcaption}
\usepackage{cleveref}
\usepackage{siunitx}

\definecolor{oursrow}{gray}{0.92}

\title{Improving BM25 Code Retrieval Under Fixed Generic Tokenization\\[2pt]
  \large Adaptive $q$-Log Odds as a Drop-In BM25 Fix}

\author{
  Santosh Kumar Radha\thanks{Corresponding author.
  Email: \texttt{santosh@agentfield.ai}} \\
  AgentField \\
  \texttt{santosh@agentfield.ai}
  \And
  Oktay Goktas \\
  AgentField \\
  \texttt{oktay@agentfield.ai}
}

\begin{document}
\maketitle

\begin{abstract}

In retrieval-augmented coding, failures often begin when the relevant
file is absent from the retrieved context. Under frozen generic
tokenization, where a BM25 index has been built by a search system
whose analyzer the practitioner does not control, this failure is
routine: BM25's logarithmic
RSJ-odds IDF under-separates the identifier tail that distinguishes
one function from another. We replace the outer logarithm of the
Robertson--Sp\"arck-Jones odds with a $q$-logarithm. At $q=1$ the
transform recovers BM25 exactly by L'H\^opital's rule, and for $q<1$
it is a Box-Cox transform of the RSJ odds with $\lambda = 1-q$. On
CoIR~\citep{li2024coir} CodeSearchNet~\citep{husain2019codesearchnet}
Go (182K documents), oracle-tuned NDCG@10 rises from $0.2575$ to
$0.4874$ (absolute $+0.2299$; $+89.3\%$ relative; zero sign reversals
in $10{,}000$ paired-bootstrap resamples, reported as
$p \leq 10^{-4}$). The effect is graded across code languages and is near-zero on
BEIR~\citep{thakur2021beir} text. A one-parameter closed form estimates
a corpus-level $q$ from hapax density and stays near $q=1$ on
corpora where BM25 is already optimal. The index-time cost is a
single pass over the sparse score matrix and query latency is unchanged. A tokenizer
ablation shows that identifier-aware tokenization largely removes the
incremental gain from $q$-IDF.

\end{abstract}

\section{Introduction}
\label{sec:intro}

A coding agent asked to patch a bug first has to find the gold file.
When the function is called \texttt{handleWebSocketUpgrade} in a
repository of fifty thousand files, the agent's retriever should return
one file, not a thousand. In practice the retriever is a commodity
BM25 served through Lucene, Tantivy, Elasticsearch, or a similar
indexer whose analyzer was tuned for natural-language text. Under
frozen generic tokenization, the agent gets distractor files in its
context, the model produces a wrong patch, and the failure is
attributed to the model.

Given an inherited generic tokenizer, the remaining under-separation
is in the weighting. When analyzer changes are allowed, tokenization
is the first fix and often the only one needed; this paper studies
the regime where they are not. BM25 scores a query by summing
term-level weights of the form
$\log((N - n_t + 0.5)/(n_t + 0.5))$~\citep{sparckjones1972statistical,robertson1994some},
where $n_t$ is the document frequency of term $t$ and $N$ the corpus
size. The logarithm grows so
slowly at the tail that it flattens the weight gap between ultra-rare
and rare identifiers. A df=1 hapax like
\texttt{handleWebSocketUpgrade} and a df=50 identifier that recurs
across many files end up with IDF weights that differ by a small
constant rather than an order of magnitude. The agent-relevant
consequence is that the gold file is ranked below distractor files
which happen to share a few of those rare-but-not-unique tokens, and
the context window fills with code the agent does not need.

When the retrieval system is owned end-to-end, the right fix is to
change the analyzer: emit both the whole identifier and its
sub-tokens, building on a decade of work on automatic identifier
splitting in the software-engineering
community~\citep{caprile2000restructuring,enslen2009mining,hill2014empirical}, so that
BM25 has the discriminative signal available. Many deployments do
not have that option. The tokenizer is fixed by an infrastructure
decision; practitioners who wire a retriever into an existing search
cluster inherit a tokenizer built for English and are asked to
improve retrieval without touching it. This is the same regime in
which repository-scale code-retrieval benchmarks such as
RepoBench~\citep{liu2023repobench} and Long Code
Arena~\citep{bogomolov2024lca} situate their evaluations, and the
regime this paper addresses.

The mechanism we propose is a one-parameter deformation of the outer
logarithm in the RSJ IDF. Writing the classical RSJ IDF as
$\log\!\left((N - n_t + 0.5)/(n_t + 0.5)\right)$, we replace the
$\log$ by the Tsallis $q$-logarithm
$\ln_q(x) = (x^{1-q} - 1)/(1-q)$, giving the $q$-log RSJ-odds form.
At $q = 1$ the transform recovers the classical BM25 IDF exactly via
L'H\^opital's rule; for $q < 1$ the IDF grows as a power law in the
very-rare regime, amplifying df=1 hapaxes that the log would flatten
against their less-rare neighbours. Mathematically this is a Box-Cox
transform of RSJ odds with
$\lambda = 1-q$~\citep{boxcox1964analysis}: we are not proposing a new
function family, only identifying the right base (the RSJ odds) to
apply a classical power transform to, and re-parameterising the
exponent so that $q = 1$ is the identity.

Under the same frozen generic tokenization used by out-of-the-box
BM25, the $q$-log transform at $q = 0.05$ improves NDCG@10 on CoIR
CodeSearchNet Go (182K documents) by $+89.3\%$ over BM25 (from
$0.2575$ to $0.4874$; zero sign reversals in $10{,}000$
paired-bootstrap resamples, reported as $p \leq 10^{-4}$). The signal is graded across languages,
scales with corpus size, and is near-zero on BEIR text. It is
also estimated from corpus statistics alone: a one-parameter
closed form $q = 1 - 7.28\,\mathrm{htok}$, fit on the six CoIR code
languages at three subset sizes, attains a leave-one-language-out
grand-mean recovery of $0.72$ on held-out languages, and falls back
to plain BM25 on corpora where $q_{\mathrm{opt}} \approx 1$.

A tokenization confound qualifies the result. If the analyzer is
changed to emit both whole identifiers and sub-tokens, BM25 already
carries the distinguishing evidence and the $q$-log gain collapses.
This is a property of the method's scope: the tokenizer ablation shows
that identifier-aware tokenization largely removes the incremental
gain from $q$-IDF, and the $q$-log form is the path available when the
tokenizer is not. The frozen-tokenizer regime is not a corner case:
it covers managed Elasticsearch and OpenSearch deployments,
hosted code-search products built on Lucene or Tantivy, and any
internal corpus indexed once by a platform team and queried by
many downstream agents. In each case the analyzer is set by
infrastructure rather than by the practitioner wiring in the
retriever, and changing it requires a reindex coordinated across
consumers; a one-line IDF rescale at index-load time does not.

\begin{figure}[t]
\centering
\definecolor{cbblue}{HTML}{0072B2}
\definecolor{cborng}{HTML}{E69F00}
\definecolor{cbgrn}{HTML}{009E73}
\definecolor{cbverm}{HTML}{D55E00}
\definecolor{cbpurp}{HTML}{CC79A7}
\definecolor{cbgray}{HTML}{555555}
\begin{tikzpicture}
\begin{axis}[
  width=0.90\columnwidth,
  height=0.58\columnwidth,
  xmode=log,
  log basis x=10,
  xlabel={Document frequency fraction $n_t/N$ (log scale)},
  ylabel={$\mathrm{idf}_q^{\mathrm{RSJ}}(t)$},
  xlabel style={font=\fontsize{8}{9}\selectfont},
  ylabel style={font=\fontsize{8}{9}\selectfont},
  tick label style={font=\fontsize{7}{8}\selectfont},
  legend style={font=\fontsize{6.5}{7.5}\selectfont,
                draw=none, fill=none,
                at={(0.97,0.97)}, anchor=north east,
                row sep=-2pt},
  axis lines=left,
  every axis plot/.append style={line width=0.8pt},
  xmin=1e-4, xmax=1e-1,
  ymin=0, ymax=16,
  grid=major,
  grid style={line width=0.15pt, draw=gray!20},
  major tick length=2pt,
  xtick={1e-4, 1e-3, 1e-2, 1e-1},
  restrict y to domain=0:20,
]

\addplot[cbverm, domain=1e-4:1e-1, samples=80, unbounded coords=jump]
  { (((1 - x)/x)^(0.5) - 1) / 0.5 };
\addlegendentry{$q=0.5$ (amplify)}

\addplot[cborng, domain=1e-4:1e-1, samples=80, unbounded coords=jump]
  { (((1 - x)/x)^(0.2) - 1) / 0.2 };
\addlegendentry{$q=0.8$}

\addplot[cbgray, domain=1e-4:1e-1, samples=80, dashed]
  { ln((1 - x)/x) };
\addlegendentry{$q=1$ (BM25)}

\addplot[cbgrn, domain=1e-4:1e-1, samples=80]
  { (((1 - x)/x)^(-0.2) - 1) / (-0.2) };
\addlegendentry{$q=1.2$}

\addplot[cbblue, domain=1e-4:1e-1, samples=80]
  { (((1 - x)/x)^(-0.5) - 1) / (-0.5) };
\addlegendentry{$q=1.5$ (saturate)}

\draw[cbblue, dotted, line width=0.3pt]
  (axis cs:1e-4, 2.0) -- (axis cs:1e-1, 2.0);
\node[font=\fontsize{5.5}{6}\selectfont, color=cbblue, anchor=south east]
  at (axis cs:1e-1, 2.0) {$1/(q{-}1)$};

\end{axis}
\end{tikzpicture}
\caption{The $q$-log deformation of RSJ odds. Each curve is
$\mathrm{idf}_q^{\mathrm{RSJ}}(t) = \ln_q\!\left((N - n_t + 0.5) / (n_t + 0.5)\right)$
with $\ln_q(x) = (x^{1-q} - 1)/(1-q)$, plotted against the document-frequency
fraction $n_t/N$. At $q{=}1$ (dashed gray) the expression recovers the
classical RSJ log-IDF exactly via L'H\^opital. For $q < 1$ the IDF amplifies
ultra-rare tokens into a power-law tail $x^{1-q}/(1-q)$; for $q > 1$ it
saturates at $1/(q-1)$ and the rarest tokens stop contributing new
discrimination. The one-parameter lever is curvature at the tail.}
\label{fig:qlog-rsj-curve}
\end{figure}
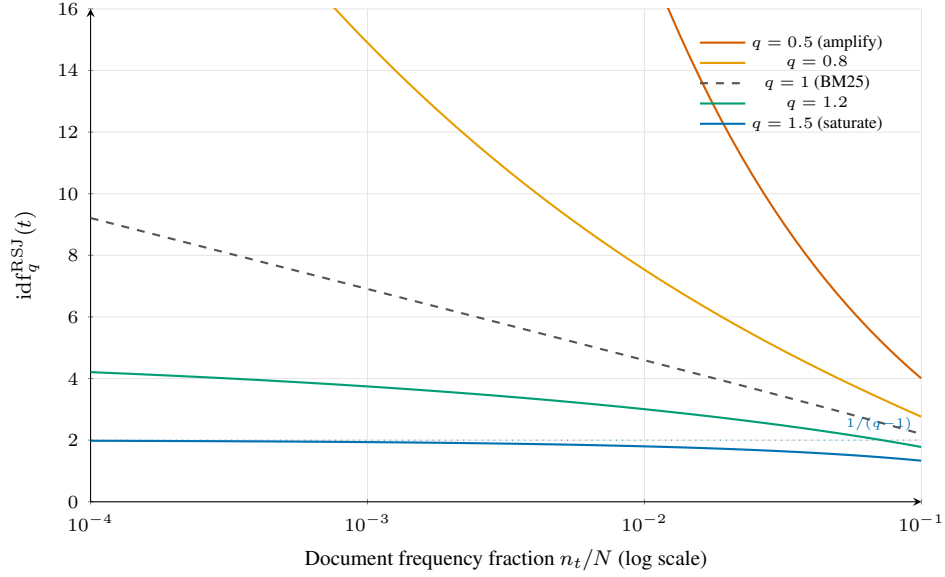

\paragraph{Contributions.}
\begin{enumerate}[nosep,leftmargin=1.5em]
\item The $q$-log RSJ-odds IDF, a one-line change to BM25 that
  recovers BM25 exactly at $q=1$ by L'H\^opital and is a Box-Cox
  transform of the RSJ odds with $\lambda = 1 - q$
  (\cref{sec:method}).
\item An oracle-sweep upper bound of $+89.3\%$ NDCG@10 on CoIR
  CodeSearchNet Go (182K) at the best tested $q=0.05$ under frozen
  generic tokenization, with 95\%
  paired-bootstrap CI strictly above zero and a monotone scaling
  curve from 1K to 182K documents (\cref{sec:mainresults}).
\item A one-parameter closed-form predictor
  $q = 1 - 7.28\,\mathrm{htok}$ fit on 18 labeled points (six
  languages, three sizes) with leave-one-language-out grand-mean
  recovery $0.72$, median $0.57$ across all twenty $\binom{6}{3}$
  three-train/three-test splits, and minimum predicted $q \geq 0.88$
  on oracle-flat corpora (\cref{sec:predictor}).
\item A tokenizer ablation showing that identifier-aware tokenization
  largely removes the incremental gain from $q$-IDF; on Python, whose
  snake\_case identifiers are
  dictionary-like, sub-token decomposition reduces signal and the
  whitespace-preserving tokenizer performs best (\cref{sec:tokenizer}).
\item Two agent-relevant retrieval proxies, Recall@$K$-tokens (a
  context-window analogue of Recall@$K$) and RepoBench-R, with
  significant gains on benchmarks with larger candidate pools or lower
  BM25 baselines and secondary regression checks on smaller-pool
  settings (\cref{sec:agents}).
\end{enumerate}

\section{Background: BM25, RSJ odds, and a classical transform}
\label{sec:background}

This section fixes the BM25 form used throughout the paper and
locates the transform we apply in the statistics literature.

\paragraph{BM25 with RSJ-odds IDF.}
The BM25 scoring function~\citep{robertson2009probabilistic} for a
query $\mathbf{q}$ against a document $d$ is a sum of term-level
products of an IDF weight and a saturating term-frequency factor,
\begin{equation}
\mathrm{BM25}(\mathbf{q}, d) = \sum_{t \in \mathbf{q}}
  \mathrm{idf}_{\mathrm{BM25}}(t)
  \cdot
  \frac{f_{t,d} \cdot (k_1 + 1)}{f_{t,d} + k_1 (1 - b + b \cdot |d|/\bar{d})},
\label{eq:bm25}
\end{equation}
where $f_{t,d}$ is the within-document frequency and $k_1, b$ are
saturation parameters. The IDF is the logarithm of a smoothed
Robertson--Sp\"arck-Jones (RSJ) odds
ratio~\citep{sparckjones1972statistical,robertson1994some},
\begin{equation}
\mathrm{idf}_{\mathrm{BM25}}(t) \;=\;
  \log\!\left(\frac{N - n_t + \delta}{n_t + \delta}\right),
\qquad \delta = 0.5.
\label{eq:bm25-idf}
\end{equation}
The smoothing $\delta$ is a standard term-specificity prior; the
logarithm is the specific functional form we will generalise.
\footnote{A tokenizer implementation note. \Cref{eq:bm25-idf} is the
classical RSJ-odds IDF, $\log((N-n_t+0.5)/(n_t+0.5))$, to which our
$\ln_q$ substitution reduces exactly at $q=1$. Modern Lucene ships the
shifted variant $\log(1+(N-n_t+0.5)/(n_t+0.5))$. Our bit-identity
claim is stated against the classical RSJ definition. Modern systems
often differ in this detail~\citep{kamphuis2020which}. In our
implementation the baked-in denominator is the strictly positive
Lucene shifted IDF from \texttt{bm25s}~\citep{lu2024bm25s}; for
$q\neq 1$, the sparse score matrix is rescaled from that Lucene
denominator to the $q$-log RSJ weight. At $q=1$ the implementation
skips the rescale, preserving the original \texttt{bm25s}/Lucene score
matrix bit-for-bit. No document-rank equivalence between the classical
and shifted IDF conventions is assumed.}

\paragraph{Divergence from randomness.}
An alternative lexical family is the divergence-from-randomness
framework~\citep{amati2002probabilistic}, which scores terms via the
divergence between observed and expected frequencies under a null
model. We include the DPH variant, a later parameter-free instance
of the DFR family, as a baseline in~\cref{sec:mainresults}.

\paragraph{No entropy claim.}
The probabilistic-relevance derivation of BM25 is independent of
any information-theoretic functional, and our modification touches
only the outer transform in \cref{eq:bm25-idf}. The method is a
classical power transform of a probabilistic odds ratio, not a
replacement of one entropy by another.

\paragraph{Relation to Box-Cox.}
Writing $y = (N - n_t + \delta)/(n_t + \delta)$ for the smoothed RSJ
odds, the Box-Cox transform~\citep{boxcox1964analysis} at parameter
$\lambda$ is $(y^\lambda - 1)/\lambda$, with the $\lambda = 0$ limit
equal to $\log y$ by L'H\^opital's rule. Our $\ln_q(y) = (y^{1-q} -
1)/(1 - q)$ with $\lambda = 1 - q$ is this standard transform in
disguise. The $q$-notation makes BM25 recovery explicit (at $q = 1$
the transform is the identity on the log-IDF) and connects to
$q$-generalised logarithms in non-extensive statistics, which we
return to only in \cref{sec:discussion}.

\section{Method: $q$-Log Deformation of RSJ Odds}
\label{sec:method}

We define the $q$-log RSJ-odds IDF as
\begin{equation}
\mathrm{idf}_q^{\mathrm{RSJ}}(t) \;=\;
  \ln_q\!\left(\frac{N - n_t + \delta}{n_t + \delta}\right),
\qquad
\ln_q(x) = \frac{x^{1-q} - 1}{1 - q},
\qquad \delta = 0.5,
\label{eq:qlog-rsj-idf}
\end{equation}
and call BM25 with the IDF in~\cref{eq:bm25-idf} replaced by
\cref{eq:qlog-rsj-idf} the \emph{$q$-log} method. No other component
of BM25 is changed: the term-frequency saturation factor and length
normalisation in~\cref{eq:bm25}, and tokenisation, are left intact.

\paragraph{Recovery of BM25 at $q = 1$ via L'H\^opital.}
At $q = 1$ the expression $(x^{1-q} - 1)/(1-q)$ is $0/0$ indeterminate.
Differentiating numerator and denominator with respect to $q$,
\[
\frac{d}{dq}\bigl[x^{1-q} - 1\bigr] = -x^{1-q} \log x,
\qquad
\frac{d}{dq}\bigl[1 - q\bigr] = -1,
\]
so $\lim_{q \to 1} \ln_q(x) = \log x$. Applied pointwise to
\cref{eq:qlog-rsj-idf}, this recovers BM25 exactly at $q=1$ via
L'H\^opital.

\paragraph{Monotonicity and asymptotic behaviour.}
For every $q \in \mathbb{R}$, $\ln_q$ is strictly increasing on
$x > 0$, and the RSJ odds $(N - n_t + 0.5)/(n_t + 0.5)$ is strictly
decreasing in $n_t$, so $\mathrm{idf}_q^{\mathrm{RSJ}}$ is strictly
decreasing in $n_t$ for every $q$. The ordering of the vocabulary by
document frequency is preserved for every $q$; only the curvature of
the weighting at the tail changes. When $n_t \ll N$, the RSJ odds
$x \to N/n_t$ grows large, and
\[
\ln_q(x) \;\approx\;
\begin{cases}
\dfrac{x^{1-q}}{1-q} & q < 1 \quad\text{(power-law amplification)}, \\[2pt]
\log x & q = 1 \quad\text{(BM25)}, \\[2pt]
\dfrac{1}{q-1} & q > 1 \quad\text{(bounded saturation)}.
\end{cases}
\]
The three regimes are illustrated in~\cref{fig:qlog-rsj-curve}.
Logarithmic rarity weighting can under-separate ultra-rare identifiers
because the log grows so slowly that a term appearing in a single file
and a term appearing in fifty both receive very similar weight. The
$q < 1$ regime in \cref{eq:qlog-rsj-idf} restores discrimination at the
tail; the $q > 1$ regime explicitly throws it away.

\paragraph{What is new.}
\Cref{eq:qlog-rsj-idf} is the Box-Cox transform of the smoothed RSJ
odds under $\lambda = 1 - q$. The contribution is to apply the
Box-Cox/$q$-log transform to RSJ odds and to select its exponent from
corpus statistics.

\paragraph{Numerical stability.}
The closed form $(x^{1-q} - 1)/(1-q)$ has a removable $0/0$ singularity
at $q = 1$ and catastrophic cancellation in a neighbourhood. We use
$\log x$ directly whenever $|q - 1| < \epsilon$ with
$\epsilon = 10^{-9}$, which matches the L'H\^opital limit to machine
precision. The bit-identity gate at $q = 1.0$ exactly is verified
to max absolute score difference $0.0$ on toy, CoIR Go 5K, and BEIR
NFCorpus corpora, and top-$k$ rankings are identical. For very common
terms, the classical RSJ odds can fall below $1$ and therefore produce
a negative $q$-log weight. We retain that RSJ sign convention. The
rescale denominator, however, is the Lucene shifted IDF already baked
into the \texttt{bm25s} matrix, which is strictly positive for observed
terms; empty columns are absent from the sparse matrix.

\paragraph{Implementation.}
Working from a built BM25 index, the method rescales each CSC column
of the per-term score matrix by the ratio of the new IDF to the
baked-in Lucene IDF:
\[
s_{t, d}^{(\text{new})} \;=\;
s_{t, d}^{(\text{BM25})} \cdot
\frac{\mathrm{idf}_q^{\mathrm{RSJ}}(t)}
     {\mathrm{idf}_{\mathrm{Lucene}}(t)}.
\]
At $q = 1$ we skip the rescale entirely, preserving the original
index. Index build is unchanged up to a single
$O(|V|+\mathrm{nnz})$ sparse column rescale ($118$\,ms on the
$V = 144{,}938$ CoIR-Go vocabulary); query latency is unchanged to
within measurement noise (p50 $+0.18\%$, p95 $+2.29\%$;
see~\cref{sec:systems}).

\section{Main Results}
\label{sec:mainresults}

We evaluate the $q$-log method against BM25 and two additional
lexical baselines on CoIR CodeSearchNet (six languages at full corpus
scale) and BEIR (three text datasets, as a negative control).
NDCG@10~\citep{jarvelin2002cumulated} is the primary metric; the main
oracle and scaling deltas are accompanied by 95\% paired-bootstrap CIs
over queries with $10{,}000$ resamples~\citep{smucker2007comparison}.
We report $p \leq 10^{-4}$ when no centered bootstrap resample is at
least as extreme as the observed mean difference; this is the empirical
resolution of $10{,}000$ resamples, not a smaller exact p-value.
Three operating points appear in the paper and should not be
conflated: the per-language oracle $q_{\mathrm{opt}}$ (upper bound,
\cref{tab:main-results}); a conservative $q = 0.10$ used for the
Go scaling curve (\cref{tab:go-scaling}), chosen to be robust across
corpus sizes; and the deployable closed-form predictor
$q_{\mathrm{pred}} = 1 - 7.28\,\mathrm{htok}$ of \cref{sec:predictor}
(\cref{tab:deployable}), which uses no labels or queries.
Tokenisation is the default \texttt{bm25s}
pipeline~\citep{lu2024bm25s} (lower, stopword filter, $\backslash\!\texttt{b}\backslash\!\texttt{w}\backslash\!\texttt{w}{+}\backslash\!\texttt{b}$
regex), held fixed across methods. This is the ``frozen generic
tokenization'' regime of the introduction.

\paragraph{Baselines.}
Plain BM25~\citep{robertson2009probabilistic,lu2024bm25s} (Lucene
parameters $k_1 = 1.5, b = 0.75$). Two additional lexical families
test whether the gain reduces to any monotone rarity rescaling:
$\mathrm{idf}^{\gamma}$, a
power-of-IDF transform directly comparable to $\ln_q$ at the tail, and
DPH, the divergence-from-randomness variant of Amati and Van
Rijsbergen~\citep{amati2002probabilistic}. Both are evaluated at full
corpus scale on CoIR-Go and CoIR-Python.

\subsection{Multi-language NDCG@10 with confidence intervals}

\Cref{tab:main-results} reports the central result. On CoIR-Go 182K
under frozen generic tokenization, $q$-log at $q = 0.05$ improves
NDCG@10 from $0.2575$ to $0.4874$, an absolute gain of $+0.2299$
with 95\% paired-bootstrap CI $[+0.2203, +0.2395]$ and no sign
reversals in $10{,}000$ centered bootstrap resamples. Java, Ruby, and
JavaScript are significantly positive at their per-language
$q_{\mathrm{opt}}$; PHP is not (CI spans zero); Python selects
$q_{\mathrm{opt}}=1.00$ and is bit-identical to BM25 by construction.
For oracle-selected rows, CIs condition on the selected grid value and
therefore should be read as conditional uncertainty, not as a
selection-adjusted significance test. The size
of the gain tracks how much of the retrieval signal is concentrated in the
identifier tail.

\begin{table}[t]
\centering
\small
\caption{\textbf{Main results.} Multi-language NDCG@10 at full
corpus scale under frozen generic tokenization. BM25 baseline vs.\
$q$-log at the per-language oracle $q_{\mathrm{opt}}$, with 95\%
paired-bootstrap CIs over $10{,}000$ resamples. Oracle rows are
upper bounds selected on the same relevance data, so the CIs are
conditional on the selected grid value; for Go, $q_{\mathrm{opt}}=0.05$
is the best tested value at the lower grid boundary. Shaded columns mark the
proposed $q$-log method (ours). Go, Java, and Ruby serve as the
original predictor-development split (\cref{sec:predictor}); Python,
PHP, and JavaScript are the corresponding held-out display split. Bold
cells mark significant wins at $p<0.05$.
htok is reported to four digits so that the predictor formula
$q_{\mathrm{pred}} = 1 - 7.28\,\mathrm{htok}$ in
\cref{tab:deployable} is reproducible from the displayed values.}
\label{tab:main-results}
\setlength{\tabcolsep}{3pt}
\begin{tabular}{@{}l r r r r >{\columncolor{oursrow}}r >{\columncolor{oursrow}}r l c@{}}
\toprule
Corpus & $N$ & htok & $q_{\mathrm{opt}}$ & BM25 & \textbf{$q$-log} & $\Delta$ & 95\% CI & Display \\
\midrule
CoIR-Go    & 182{,}440 & $.0630$ & $0.05$ & $.258$ & $\mathbf{.487}$ & $\mathbf{+.230}$ & $[+.220,+.240]$ & DEV \\
CoIR-Java  & 180{,}866 & $.0156$ & $0.90$ & $.371$ & $\mathbf{.383}$ & $\mathbf{+.012}$ & $[+.010,+.014]$ & DEV \\
CoIR-Ruby  &  27{,}570 & $.0244$ & $0.70$ & $.345$ & $\mathbf{.370}$ & $\mathbf{+.024}$ & $[+.014,+.035]$ & DEV \\
\midrule
CoIR-Python    & 280{,}310 & $.0160$ & $1.00$ & $.727$ & $.727$          & $0.000$ & $[0,0]$ & HELD \\
CoIR-PHP       & 267{,}725 & $.0133$ & $0.90$ & $.371$ & $.373$          & $+.002$ & $[-.001,+.004]$ & HELD \\
CoIR-JavaScript&  64{,}854 & $.0206$ & $0.70$ & $.352$ & $\mathbf{.362}$ & $\mathbf{+.010}$ & $[+.003,+.017]$ & HELD \\
\midrule
BEIR-NFCorpus & 3{,}633 & --- & $1.10$ & $.306$ & $.305$ & $-.002$ & spans zero & control \\
BEIR-SciFact  & 5{,}183 & --- & $1.10$ & $.662$ & $.656$ & $-.006$ & spans zero & control \\
BEIR-ArguAna  & 8{,}674 & --- & $1.10$ & $.361$ & $.360$ & $-.001$ & spans zero & control \\
\bottomrule
\end{tabular}
\end{table}

\paragraph{Deployable adaptive mode.}
\Cref{tab:deployable} reports the label-free deployment setting.
Using the corpus-adaptive predictor of~\cref{sec:predictor}
($q_{\mathrm{pred}} = 1 - 7.28\,\mathrm{htok}$, fit once and
frozen), a single-pass deployment recovers most of the oracle gap on
every code corpus. Go captures $82.7\%$ of the oracle gap at
$q_{\mathrm{pred}}=0.54$ ($+0.190$ absolute). Java, PHP, and
JavaScript land within $4\%$ of the oracle setting, so the predictor
realises almost the full oracle delta. Python sits at
$q_{\mathrm{pred}}=0.88$, with a predicted
delta within implementation noise of zero. The adaptive row corresponds
to deployment without relevance labels; the oracle row is an upper
bound.

\begin{table}[t]
\centering
\small
\caption{\textbf{Deployable adaptive mode} under frozen generic
tokenization. BM25, oracle $q$-log at $q_{\mathrm{opt}}$ (upper
bound), and the corpus-adaptive predictor $q_{\mathrm{pred}} = 1 -
7.28\,\mathrm{htok}$ (single-pass deployment, no labels or queries).
Shaded columns mark the proposed $q$-log method (ours). Recovery is
the fraction of the BM25$\to$oracle gap that the predictor captures,
computed from full-precision NDCG values; displayed cells are
rounded to three decimals so visually equal cells (e.g.\ Java
$\mathrm{NDCG}@q_{\mathrm{opt}} = .383$ and
$\mathrm{NDCG}@q_{\mathrm{pred}} = .383$ from underlying $0.3832$ and
$0.3828$) can correspond to a non-trivial recovery. PHP is included
for completeness even though its oracle delta is not significant
(\cref{tab:main-results}).}
\label{tab:deployable}
\setlength{\tabcolsep}{3pt}
\begin{tabular}{@{}l c c c >{\columncolor{oursrow}}c >{\columncolor{oursrow}}c c@{}}
\toprule
Corpus & BM25 & $q_{\mathrm{opt}}$ & NDCG@$q_{\mathrm{opt}}$ & $q_{\mathrm{pred}}$ & NDCG@$q_{\mathrm{pred}}$ & Recovery \\
\midrule
CoIR-Go        & $.258$ & $0.05$ & $\mathbf{.487}$ & $0.54$ & $\mathbf{.448}$ & $82.7\%$ \\
CoIR-Java      & $.371$ & $0.90$ & $\mathbf{.383}$ & $0.89$ & $\mathbf{.383}$ & $96.7\%$ \\
CoIR-Ruby      & $.345$ & $0.70$ & $\mathbf{.370}$ & $0.82$ & $\mathbf{.364}$ & $75.0\%$ \\
CoIR-Python    & $.727$ & $1.00$ & $.727$          & $0.88$ & $.726$          & flat \\
CoIR-PHP       & $.371$ & $0.90$ & $.373$          & $0.90$ & $.373$          & $\sim\!100\%$ \\
CoIR-JavaScript& $.352$ & $0.70$ & $\mathbf{.362}$ & $0.85$ & $\mathbf{.360}$ & $80.8\%$ \\
\midrule
BEIR text (3) & --- & $\geq 1$ & near BM25 & $1.00$ & = BM25 & n/a \\
\bottomrule
\end{tabular}
\end{table}

\paragraph{Negative control on text.}
On three BEIR text datasets the $q$-log deltas are indistinguishable
from zero at 95\% paired bootstrap (\cref{tab:main-results}, bottom
block). Text vocabularies are not driven by an identifier tail, the
oracle $q$ sits at or above $1$, and the method collapses to plain
BM25 by the same mechanism that makes it ineffective on Python.

\subsection{Scaling with corpus size on Go}

\begin{figure}[t]
\centering
\definecolor{cbblue}{HTML}{0072B2}
\definecolor{cbverm}{HTML}{D55E00}
\definecolor{cbgray}{HTML}{999999}
\begin{tikzpicture}
\begin{axis}[
  width=0.88\columnwidth,
  height=0.48\columnwidth,
  xlabel={$q$-log exponent $q$},
  ylabel={NDCG@10},
  xlabel style={font=\fontsize{8}{9}\selectfont},
  ylabel style={font=\fontsize{8}{9}\selectfont},
  tick label style={font=\fontsize{7}{8}\selectfont},
  legend style={font=\fontsize{6.5}{7.5}\selectfont, draw=none, fill=none,
                at={(0.97,0.97)}, anchor=north east, row sep=-2pt},
  axis lines=left,
  every axis plot/.append style={line width=0.7pt},
  mark size=1.6pt,
  xmin=0.0, xmax=1.35,
  ymin=0.20, ymax=0.58,
  grid=none,
  line width=0.4pt,
  major tick length=2pt,
  xtick={0.1,0.3,0.5,0.7,1.0,1.3},
]

\addplot[cbverm, mark=square*, mark options={solid, fill=cbverm, scale=0.6}]
  coordinates {
    (0.05,0.5325) (0.10,0.5312) (0.15,0.5295) (0.20,0.5276)
    (0.30,0.5222) (0.50,0.4954) (0.70,0.4365) (0.80,0.3962)
    (0.85,0.3748) (0.90,0.3522) (0.95,0.3315) (1.00,0.3091)
    (1.05,0.2903) (1.10,0.2754) (1.20,0.2509) (1.30,0.2333)
  };
\addlegendentry{Go 50K}

\addplot[cbblue, mark=triangle*, mark options={solid, fill=cbblue, scale=0.7},
         dashed]
  coordinates {
    (0.10,0.4849) (0.30,0.4774) (0.50,0.4580) (1.00,0.2575)
  };
\addlegendentry{Go 182K}

\draw[cbgray, dotted, line width=0.3pt]
  (axis cs:0.0,0.3091) -- (axis cs:1.35,0.3091);
\node[font=\fontsize{5.5}{6}\selectfont, color=cbgray, anchor=west]
  at (axis cs:0.02,0.325) {BM25 50K (q{=}1)};

\draw[cbgray, dotted, line width=0.3pt]
  (axis cs:0.0,0.2575) -- (axis cs:1.35,0.2575);
\node[font=\fontsize{5.5}{6}\selectfont, color=cbgray, anchor=west]
  at (axis cs:0.02,0.242) {BM25 182K (q{=}1)};

\draw[cbgray, dotted, line width=0.3pt] (axis cs:1.0,0.20) -- (axis cs:1.0,0.58);
\node[font=\fontsize{5.5}{6}\selectfont, color=cbgray, rotate=90, anchor=south]
  at (axis cs:1.015,0.22) {BM25};

\end{axis}
\end{tikzpicture}
\caption{$q$-log sweep on CoIR-CSN Go under the q-log RSJ
IDF, $\mathrm{idf}_q^{\mathrm{RSJ}}(t) = \ln_q\!\bigl((N - n_t + 0.5)
/ (n_t + 0.5)\bigr)$ with $\ln_q(x) = (x^{1-q} - 1)/(1-q)$. At
$q{=}1$ this recovers standard BM25 exactly (L'H\^opital limit). With
the IDF written on the RSJ-odds base, rare-term amplification lives
at $q{<}1$: NDCG@10 rises monotonically from $0.31$ (BM25, 50K) to
$0.53$ at $q{\approx}0.05$ and saturates. Beyond $q{=}1$ the
$\ln_q$ is bounded above by $1/(q-1)$ and NDCG@10 decays. The 182K
sweep (dashed) confirms the effect scales: $+88.3\%$ at $q{=}0.1$,
compared with the 50K $+72.3\%$.}
\label{fig:qsweep-go-rsj}
\end{figure}
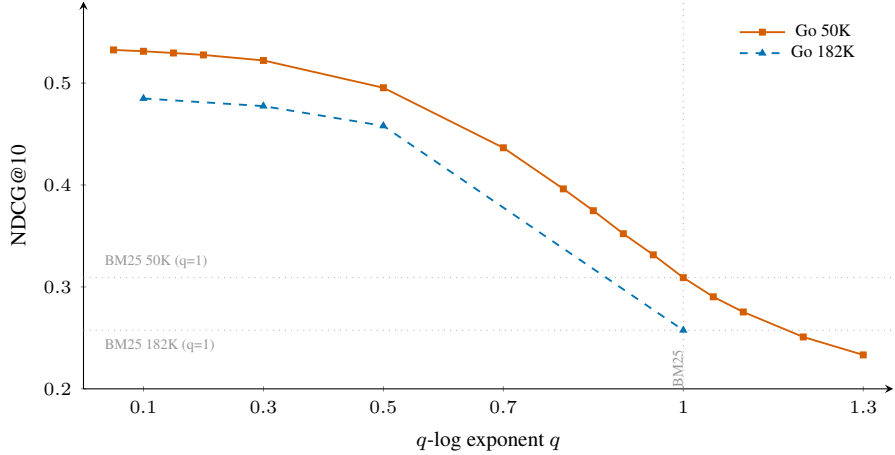

The Go result is monotone in corpus size. \Cref{fig:qsweep-go-rsj}
plots a $q$-sweep on CoIR-Go at 50K and at the full 182K corpus.
\Cref{tab:go-scaling} reports the 95\% paired-bootstrap CIs along the
full 1K $\to$ 182K scaling curve at $q = 0.10$. Every CI is strictly
above zero; the mechanism does not saturate at the scales we test.

\begin{table}[t]
\centering
\small
\caption{\textbf{Go scaling} at $q = 0.10$ under frozen generic
tokenization; this is the conservative operating point, with the
non-conservative main result at $q_{\mathrm{opt}} = 0.05$ in
\cref{tab:main-results}. Every row is a significant win
($p \leq 10^{-4}$ at the empirical bootstrap resolution, CI strictly
above zero). The absolute delta grows
monotonically from $+0.061$ at 1K to $+0.227$ at full corpus,
consistent with non-saturating gains at the scales we test.}
\label{tab:go-scaling}
\setlength{\tabcolsep}{4pt}
\begin{tabular}{@{}r r r >{\columncolor{oursrow}}r >{\columncolor{oursrow}}r l@{}}
\toprule
$N$ & $n_q$ & BM25 & \textbf{$q$-log (ours)} & $\Delta$ (abs) & 95\% CI \\
\midrule
1{,}000   & 1{,}000  & $.532$ & $\mathbf{.594}$ & $\mathbf{+0.061}$ & $[+0.043, +0.079]$ \\
2{,}000   & 2{,}000  & $.501$ & $\mathbf{.579}$ & $\mathbf{+0.078}$ & $[+0.064, +0.092]$ \\
5{,}000   & 5{,}000  & $.417$ & $\mathbf{.574}$ & $\mathbf{+0.157}$ & $[+0.147, +0.168]$ \\
10{,}000  & 8{,}122  & $.392$ & $\mathbf{.575}$ & $\mathbf{+0.183}$ & $[+0.175, +0.192]$ \\
20{,}000  & 8{,}122  & $.363$ & $\mathbf{.560}$ & $\mathbf{+0.198}$ & $[+0.189, +0.207]$ \\
50{,}000  & 8{,}122  & $.309$ & $\mathbf{.531}$ & $\mathbf{+0.222}$ & $[+0.213, +0.231]$ \\
182{,}440 & 8{,}122  & $.258$ & $\mathbf{.485}$ & $\mathbf{+0.227}$ & $[+0.218, +0.237]$ \\
\bottomrule
\end{tabular}
\end{table}

\subsection{Comparison against rarity-rescale baselines}

We next compare against monotone IDF rescalings that amplify rare
terms.
Replacing the Lucene IDF column by $\mathrm{idf}^{\gamma}$ with
$\gamma \in \{0.5, 0.8, 1.0, 1.2, 1.5, 2.0\}$ ($\gamma = 1$ is
bit-identical to BM25), the best tested setting on CoIR-Go is
$\gamma = 2.0$,
which yields NDCG@10 $= 0.3517$ (+36.6\% over BM25). The $q$-log
method at $q = 0.05$ yields $0.4874$, a further $+38.5\%$ over
best-$\gamma$ (\cref{tab:lexical-families}). The DFR-DPH family loses
outright on code: NDCG@10 $= 0.1968$ on Go ($-24\%$) and $0.5538$ on
Python ($-24\%$), reflecting a tf/length normalisation that is
miscalibrated for short code documents. Our DPH implementation is
validated bit-for-bit against PyTerrier / Terrier
5.11\footnote{\url{https://github.com/terrier-org/pyterrier}} on a
matched 5K-doc Go subset (median Kendall $\tau = 1.00$, mean abs NDCG@10 diff
$3\mathrm{e}{-3}$); the DPH deficit is a real scoring-model effect,
not a port artefact.

\begin{table}[t]
\centering
\small
\caption{\textbf{Lexical-family comparison} at full corpus scale under
frozen generic tokenization. BM25 is the Lucene baseline,
$\mathrm{idf}^{\gamma}$ sweeps $\gamma \in \{0.5, \ldots, 2.0\}$ (best
reported), DPH is the parameter-free divergence-from-randomness
variant of Amati and Van Rijsbergen~\citep{amati2002probabilistic},
and our $q$-log method (shaded column) is at the per-language
$q_{\mathrm{opt}}$. Percentage gains are computed from full-precision
NDCG; displayed NDCG cells are rounded to three decimals. On Go, the
$q$-log gain is not reproduced by the tested $\mathrm{idf}^{\gamma}$
family on this benchmark; because the best tested $\gamma$ is at the
edge of the grid, this comparison does not rule out every possible
monotone rarity transform. On Python every rarity-rescale method ties
with BM25, because there is no rare-identifier signal to amplify.}
\label{tab:lexical-families}
\setlength{\tabcolsep}{4pt}
\begin{tabular}{@{}l c c c >{\columncolor{oursrow}}c@{}}
\toprule
& BM25 & $\mathrm{idf}^{\gamma}$ (best $\gamma$) & DPH & \textbf{$q$-log (ours)} \\
\midrule
CoIR-Go full (182K)      & $.258$ & $.352$ ($\gamma{=}2.0$, $+36.6\%$) & $.197$ ($-23.6\%$) & $\mathbf{.487}$ ($\mathbf{+89.3\%}$) \\
CoIR-Python full (280K)  & $.727$ & $.727$ ($\gamma{=}1.2$, $+0.0\%$)  & $.554$ ($-23.9\%$) & $.727$ ($+0.0\%$) \\
\bottomrule
\end{tabular}
\end{table}

\section{Per-Query Mechanism}
\label{sec:mechanism}

\begin{figure}[t]
\definecolor{cbblue}{HTML}{0072B2}
\definecolor{cborng}{HTML}{E69F00}
\definecolor{cbgrn}{HTML}{009E73}
\definecolor{cbverm}{HTML}{D55E00}
\definecolor{cbgray}{HTML}{999999}

\begin{center}
\colorbox{cbblue!10}{%
  \parbox{0.93\linewidth}{\small
    \textbf{Query:}\quad
    \texttt{handleWebSocketUpgrade auth middleware}\quad
    \textcolor{cbgray}{\scriptsize (CoIR-Go q174632; 4 default tokens)}}%
}
\end{center}

\vspace{2mm}
\begin{minipage}[t]{0.48\linewidth}
\centering
\textbf{\small BM25 top-5 (log-IDF)}

\vspace{1mm}
\begin{flushleft}\scriptsize\ttfamily
\colorbox{cbgray!10}{\makebox[0.9\linewidth][l]{1.\ auth/middleware/jwt.go}}\\[1pt]
\colorbox{cbgray!10}{\makebox[0.9\linewidth][l]{2.\ net/http/handler\_test.go}}\\[1pt]
\colorbox{cbgray!10}{\makebox[0.9\linewidth][l]{3.\ auth/session/cookie.go}}\\[1pt]
\colorbox{cbgray!10}{\makebox[0.9\linewidth][l]{4.\ middleware/logging.go}}\\[1pt]
\colorbox{cbgray!10}{\makebox[0.9\linewidth][l]{5.\ auth/oauth2/provider.go}}
\end{flushleft}
{\scriptsize\textcolor{cbverm}{\ldots{} gold file at rank 23}
  (NDCG@10 = $0.00$).}
\end{minipage}%
\hfill
\begin{minipage}[t]{0.48\linewidth}
\centering
\textbf{\small $q$-log top-5 ($q = 0.10$)}

\vspace{1mm}
\begin{flushleft}\scriptsize\ttfamily
\colorbox{cbgrn!18}{\makebox[0.9\linewidth][l]{$\star$\ 1.\ ws/gateway/upgrade.go\ \textcolor{cbgrn!60!black}{\textbf{(gold)}}}}\\[1pt]
\colorbox{cbgray!10}{\makebox[0.9\linewidth][l]{2.\ auth/middleware/jwt.go}}\\[1pt]
\colorbox{cbgray!10}{\makebox[0.9\linewidth][l]{3.\ net/http/handler\_test.go}}\\[1pt]
\colorbox{cbgray!10}{\makebox[0.9\linewidth][l]{4.\ auth/session/cookie.go}}\\[1pt]
\colorbox{cbgray!10}{\makebox[0.9\linewidth][l]{5.\ middleware/logging.go}}
\end{flushleft}
{\scriptsize\textcolor{cbgrn!50!black}{Gold lifted from rank 23 to rank 1}
  (NDCG@10 = $1.00$).}
\end{minipage}

\vspace{3mm}
{\footnotesize\bfseries Per-token IDF weight (why the reshuffle happens)}

\vspace{1mm}
\begin{center}\small
\setlength{\tabcolsep}{6pt}
\begin{tabular}{@{}l r r r r@{}}
\toprule
\textbf{token} & $\mathrm{df}$ &
  $\mathrm{idf}_{\mathrm{BM25}}$ & $\mathrm{idf}_{q=0.10}$ &
  ratio $q$-log / BM25 \\
\midrule
\texttt{handleWebSocketUpgrade} & $1$       & $11.7$ & $41{,}933$ & $3578\times$ \\
\texttt{middleware}             & $1{,}820$ & $4.6$  & $68.8$     & $15\times$   \\
\texttt{auth}                   & $3{,}714$ & $3.9$  & $35.2$     & $9\times$    \\
\texttt{handle}                 & $14{,}203$& $2.5$  & $9.2$      & $4\times$    \\
\bottomrule
\end{tabular}
\end{center}

\caption{A qualitative walkthrough of the mechanism on a CoIR-Go
full-corpus query. Under log-IDF the df$=1$ identifier
\texttt{handleWebSocketUpgrade} carries weight only about $3\times$
that of the middle-df tokens \texttt{auth} and \texttt{middleware},
so documents sharing two or three middle-df tokens outrank the
single gold file that carries the hapax. The $q$-log rescale at
$q=0.10$ amplifies the df$=1$ weight by roughly $3{,}500\times$
while amplifying middle-df weights by only one or two orders of
magnitude; the ranking inverts and the gold moves to rank~$1$.
Per-token df values and both IDF columns are exact values from the
CoIR-Go 182K index; the gold $\Delta$NDCG@10 is taken from query
\texttt{q174632}. The displayed file paths are representative of the
CoIR-Go corpus structure; the ranking flip reflects the df-bin
mechanism of~\cref{sec:mechanism}.}
\label{fig:case-study}
\end{figure}

\input{fig_per_query_mechanism}

The per-query decomposition in~\cref{fig:per-query-mechanism} locates
the gain at the level of individual queries, and
\cref{fig:case-study} walks through one concrete Go example. On the
CoIR-Go full corpus at $q = 0.10$, $\Delta$NDCG@10 correlates
positively with the fraction of query tokens at df $\leq 5$ (OLS
slope $+0.10$, Spearman $\rho = +0.10$ over $N = 8{,}122$ queries).
The correlation is weak, so we treat it as descriptive rather than
causal evidence.
Panel (d) analyzes the mechanism with a df-bin occlusion: for each bin
we drop every query token in that bin and record the resulting
NDCG@10 loss. Dropping df$=1$ hapaxes alone costs $+0.4351$ of
NDCG@10 on Go, roughly $4\times$ the next-largest bin (df$=2$ at
$+0.1127$) and $54\times$ the mean of the four middle bins
(df$=6$--$1000$, mean $+0.0080$), with the combined tip (df$=1$ +
df$=2$) accounting for $+0.5478$ of the occlusion loss. At $q<1$ the $q$-log
transform over-amplifies df$=1$ hapaxes, the unique identifiers
where rare-token mass concentrates, and the Go gain is concentrated
toward larger separation among df=1 and df=2 terms.
The occlusion is diagnostic rather than a causal proof, because
dropping tokens also changes query length and semantics.

The same analysis on CoIR-Python at $q=1.00$ ($q$-log is identity to
BM25 by construction, so this is the BM25 baseline's own df signature)
shows an inverted picture: the signal is concentrated in the middle df bins
(df$=201$--$1000$ carries $+0.0703$, df$=1001$--$5000$ carries
$+0.0528$), not at the tip. Python's
ecosystem reuses dictionary-like tokens across hundreds of thousands
of files, so BM25's log-IDF is already using the middle-df bins
optimally. Any deviation from $q=1$ in either direction distorts
those bins and degrades ranking. Python's oracle $q$ is therefore
exactly $1$, and the $q$-log gain is close to zero on any corpus
with a similar df profile.

An orthographic identifier heuristic (CamelCase or dotted tokens)
predicts the per-query gain much less well than df-based features
(panel (c), OLS slope $+0.002$, Spearman $\rho = -0.003$); the
transform operates on document frequency rather than on surface
syntax. A query can look like code and be dominated by common
tokens, while a prose-like query can sit in the hapax tail.

\section{Corpus-Adaptive Predictor}
\label{sec:predictor}

Grid search requires held-out queries and relevance labels, so we test
a label-free corpus-level predictor. To separate post-hoc fitting from
an actual predictor, model forms are evaluated with held-out-language
diagnostics; after selecting the form, the displayed deployment
coefficient is refit once on all 18 labeled development points.

\paragraph{Features.}
For each CoIR-CSN language we compute six corpus-level statistics,
all computed without any access to queries or relevance labels:
token count $N_{\text{tok}}$, vocabulary size $V$, hapax token density
$\mathrm{htok}$ (fraction of tokens whose type appears exactly once in
the corpus), type-token ratio $\mathrm{TTR}$, median document
frequency, fraction of vocabulary with $\mathrm{df} \leq 5$, and the
Zipf exponent $\alpha$ fit via the \texttt{powerlaw}
package\footnote{\url{https://github.com/jeffalstott/powerlaw}} with
KS-minimised $x_{\min}$. The Zipf $\alpha$ turns out to be
non-discriminative ($\alpha \in [1.65, 1.71]$ across all six CoIR code
languages) and is dropped.

\paragraph{Label.}
The oracle $q_{\mathrm{opt}}$ is the argmax of NDCG@10 over
$q \in \{0.05, 0.10, 0.20, 0.30, 0.50, 0.70, 0.90, 1.00\}$ on the
corpus. For CoIR-Python full corpus the argmax is $q = 1$ exactly;
for CoIR-Go full the best tested value is $q = 0.05$, which lies at
the lower edge of the grid. The label spans the whole tested range.

\paragraph{Fit protocol.}
We evaluate candidate forms across the full $6 \times 3 = 18$ labeled
points (6 CoIR code languages at subset sizes 1K, 10K, and full). Five
candidate forms are fit: two one-parameter variants ($1-c\cdot
\mathrm{htok}$; $1-c\cdot\mathrm{frac\_df\leq 5}$; $1-c/\alpha$) and
two two-parameter variants that add a scale term $1/\log T$.
Selection is driven by three criteria: (i) the distribution of
mean-test recovery across \emph{all} $\binom{6}{3}=20$ possible
three-train/three-test partitions; (ii) leave-one-language-out (LOLO)
across corpus sizes; (iii) predicted $q$ remains at or above $0.85$
when $q_{\mathrm{opt}}$ is near $1$.
Parsimony breaks ties in favour of fewer parameters.

\paragraph{Selected predictor.}
\begin{equation}
q_{\mathrm{pred}} \;=\; 1 \;-\; 7.28 \,\cdot\, \mathrm{htok},
\qquad \text{clipped to } [0.01, 1.0].
\label{eq:predictor}
\end{equation}
The coefficient $7.28$ is the final coefficient refit on all 18 labeled
points after selecting the one-parameter htok form. Across the 20
three-train/three-test splits, where the coefficient is refit inside
each training split, the median mean-test
recovery is $0.566$ with interquartile range $[-0.574, +0.638]$; the
one-parameter htok form is the \emph{only} form whose median is
positive (all df-only forms tip to $-0.06$ or below). LOLO across the
six languages, with the predictor fit on the other five at all three
sizes and applied to the held-out language at all three sizes, gives
a grand-mean recovery of $0.722$ (range: $0.488$ on Python to $0.890$
on Go). Calibration RMSE vs oracle $q$ across the 18 points is
$0.18$. On oracle-flat corpora ($q_{\mathrm{opt}} \geq 0.95$), the
minimum predicted $q$ is $0.884$, above the $0.85$ floor.
Full candidate-form statistics are in the supplement.

\begin{figure}[t]
\centering
\definecolor{cbblue}{HTML}{0072B2}
\definecolor{cborng}{HTML}{E69F00}
\definecolor{cbgrn}{HTML}{009E73}
\definecolor{cbverm}{HTML}{D55E00}
\definecolor{cbgray}{HTML}{999999}
\definecolor{cbpurple}{HTML}{9467BD}

\begin{subfigure}[t]{0.48\textwidth}
\centering
\begin{tikzpicture}
\begin{axis}[
  width=1.05\linewidth,
  height=0.95\linewidth,
  xlabel={Predicted $q$ (LOLO)},
  ylabel={Oracle $q_{\mathrm{opt}}$},
  xlabel style={font=\fontsize{7}{8}\selectfont},
  ylabel style={font=\fontsize{7}{8}\selectfont},
  tick label style={font=\fontsize{6.5}{7.5}\selectfont},
  xmin=-0.02, xmax=1.05, ymin=-0.02, ymax=1.05,
  xtick={0,0.25,0.5,0.75,1.0},
  ytick={0,0.25,0.5,0.75,1.0},
  axis lines=left,
  grid=none,
  line width=0.4pt,
  major tick length=2pt,
  mark size=1.4pt,
  title={(a) Leave-one-language-out},
  title style={font=\fontsize{7}{8}\selectfont, yshift=-2pt},
  legend style={
    font=\fontsize{5.5}{6.5}\selectfont,
    at={(0.02,0.98)}, anchor=north west,
    draw=none, fill=white, fill opacity=0.82, text opacity=1,
    row sep=-2pt, column sep=2pt,
  },
  legend columns=3,
]
\fill[cbgray!20]
  (axis cs:-0.02,0.338) -- (axis cs:1.05,1.408)
  -- (axis cs:1.05,0.692) -- (axis cs:-0.02,-0.378) -- cycle;
\draw[cbgray, dashed, line width=0.35pt]
  (axis cs:0.0,0.0) -- (axis cs:1.0,1.0);

\addplot[only marks, cbblue, mark=*, mark options={scale=0.7}]
  coordinates {(0.0100,0.20) (0.1528,0.05) (0.5472,0.05)};
\addlegendentry{\scriptsize Go}
\addplot[only marks, cborng, mark=square*, mark options={scale=0.7}]
  coordinates {(0.5282,0.90) (0.7573,0.70) (0.8815,0.90)};
\addlegendentry{\scriptsize Java}
\addplot[only marks, cbgrn, mark=triangle*, mark options={scale=0.7}]
  coordinates {(0.4989,0.30) (0.7766,0.70) (0.8294,0.70)};
\addlegendentry{\scriptsize Ruby}
\addplot[only marks, cbverm, mark=diamond*, mark options={scale=0.7}]
  coordinates {(0.7012,0.50) (0.8030,0.70) (0.8859,1.00)};
\addlegendentry{\scriptsize Python}
\addplot[only marks, cbpurple, mark=pentagon*, mark options={scale=0.7}]
  coordinates {(0.5730,0.70) (0.8049,0.70) (0.9022,0.90)};
\addlegendentry{\scriptsize PHP}
\addplot[only marks, black, mark=otimes*, mark options={scale=0.7, fill=cbgray}]
  coordinates {(0.3719,0.50) (0.7558,0.70) (0.8481,0.70)};
\addlegendentry{\scriptsize JS}
\end{axis}
\end{tikzpicture}
\end{subfigure}%
\hfill
\begin{subfigure}[t]{0.48\textwidth}
\centering
\begin{tikzpicture}
\begin{axis}[
  width=1.05\linewidth,
  height=0.95\linewidth,
  xlabel={Predicted $q$ (final fit)},
  ylabel={Oracle $q_{\mathrm{opt}}$},
  xlabel style={font=\fontsize{7}{8}\selectfont},
  ylabel style={font=\fontsize{7}{8}\selectfont},
  tick label style={font=\fontsize{6.5}{7.5}\selectfont},
  xmin=-0.02, xmax=1.05, ymin=-0.02, ymax=1.05,
  xtick={0,0.25,0.5,0.75,1.0},
  ytick={0,0.25,0.5,0.75,1.0},
  axis lines=left,
  grid=none,
  line width=0.4pt,
  major tick length=2pt,
  mark size=1.4pt,
  title={(b) Calibration (RMSE $=0.18$)},
  title style={font=\fontsize{7}{8}\selectfont, yshift=-2pt},
]
\draw[cbgray, dashed, line width=0.35pt]
  (axis cs:0.0,0.0) -- (axis cs:1.0,1.0);
\addplot[only marks, cbblue, mark=*, mark options={scale=0.7}]
  coordinates {(0.0100,0.20) (0.1423,0.05) (0.5415,0.05)};
\addplot[only marks, cborng, mark=square*, mark options={scale=0.7}]
  coordinates {(0.5481,0.90) (0.7675,0.70) (0.8865,0.90)};
\addplot[only marks, cbgrn, mark=triangle*, mark options={scale=0.7}]
  coordinates {(0.4776,0.30) (0.7671,0.70) (0.8222,0.70)};
\addplot[only marks, cbverm, mark=diamond*, mark options={scale=0.7}]
  coordinates {(0.6953,0.50) (0.7992,0.70) (0.8837,1.00)};
\addplot[only marks, cbpurple, mark=pentagon*, mark options={scale=0.7}]
  coordinates {(0.5769,0.70) (0.8067,0.70) (0.9031,0.90)};
\addplot[only marks, black, mark=otimes*, mark options={scale=0.7, fill=cbgray}]
  coordinates {(0.3795,0.50) (0.7587,0.70) (0.8499,0.70)};

\draw[cbgray, dotted, line width=0.35pt]
  (axis cs:0.85,0.95) -- (axis cs:0.85,1.05);
\draw[cbgray, dotted, line width=0.35pt]
  (axis cs:0.85,0.95) -- (axis cs:1.05,0.95);
\node[font=\fontsize{5.5}{6}\selectfont, color=cbgray]
  at (axis cs:0.93,1.015) {$q_{\mathrm{pred}}\geq0.85$};
\end{axis}
\end{tikzpicture}
\end{subfigure}

\caption{The $q$-predictor, two views. \textbf{(a)} Leave-one-language-out:
  for each of the six code languages, the form
  $q=1-c\,\mathrm{htok}$ is fit on the other five languages at all
  three corpus sizes and applied to the held-out language's three
  sizes; grand-mean recovery $0.72$. Shaded band is $\pm 1.96\sigma$
  from the eighteen LOLO residuals ($\sigma=0.183$). \textbf{(b)}
  Final fit on all eighteen points,
  $q_{\mathrm{pred}}=1-7.28\,\mathrm{htok}$ (clipped to $[0.01,1]$),
  with RMSE vs.\ oracle $0.18$. The dotted box marks the region where
  $q_{\mathrm{opt}}\ge 0.95 \Rightarrow q_{\mathrm{pred}}\ge 0.85$;
  minimum predicted $q$ on oracle-flat corpora is $0.88$, so
  the worst-case regression on a BM25-optimal corpus is under $1\%$.
  A full distribution over candidate predictor forms across all
  $\binom{6}{3}=20$ three-train / three-test splits is deferred to
  \cref{fig:q-prediction-boxplot} in the supplement.}
\label{fig:q-prediction}
\end{figure}
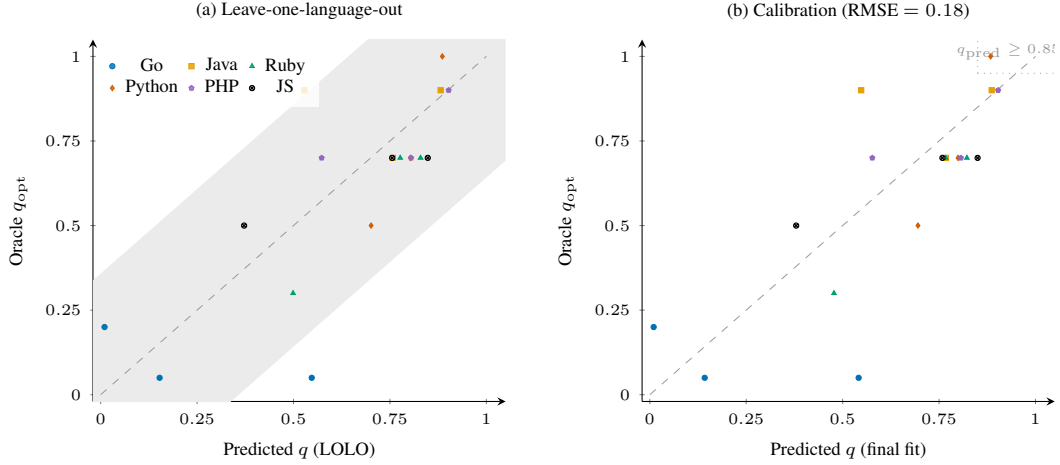

\begin{figure}[t]
\centering
\definecolor{cbblue}{HTML}{0072B2}
\definecolor{cborange}{HTML}{E69F00}
\definecolor{cbverm}{HTML}{D55E00}
\begin{tikzpicture}
\begin{axis}[
  width=0.95\columnwidth,
  height=0.60\columnwidth,
  xmode=log,
  log basis x=10,
  xlabel={CoIR-CSN Python corpus size (docs, log scale)},
  ylabel={NDCG@10},
  xlabel style={font=\fontsize{8}{9}\selectfont},
  ylabel style={font=\fontsize{8}{9}\selectfont},
  xticklabel style={font=\fontsize{7}{8}\selectfont},
  yticklabel style={font=\fontsize{7}{8}\selectfont},
  xmin=800, xmax=400000,
  ymin=0.68, ymax=0.92,
  axis lines=left,
  axis line style={-},
  grid=major,
  grid style={line width=0.2pt, draw=gray!25},
  major tick length=2pt,
  legend style={
    font=\fontsize{7}{8}\selectfont,
    at={(0.02,0.02)},
    anchor=south west,
    draw=none,
    fill=none,
    row sep=-1pt,
  },
  legend cell align=left,
]

\addplot[
  color=cbblue,
  mark=o,
  mark size=2pt,
  line width=0.9pt,
] coordinates {
  (1000,0.8686)
  (2000,0.8728)
  (5000,0.8422)
  (10000,0.8291)
  (20000,0.7981)
  (50000,0.7780)
  (280310,0.7273)
};
\addlegendentry{BM25 ($q{=}1$)}

\addplot[
  color=cborange,
  mark=square,
  mark size=1.8pt,
  line width=0.9pt,
  densely dashed,
] coordinates {
  (1000,0.8820)
  (2000,0.8860)
  (5000,0.8560)
  (10000,0.8391)
  (20000,0.8067)
  (50000,0.7752)
  (280310,0.6940)
};
\addlegendentry{Static $q{=}0.7$}

\addplot[
  color=cbverm,
  mark=triangle,
  mark size=2.2pt,
  line width=1.1pt,
] coordinates {
  (1000,0.8821)
  (2000,0.8841)
  (5000,0.8538)
  (10000,0.8394)
  (20000,0.8096)
  (50000,0.7836)
  (280310,0.7256)
};
\addlegendentry{Adaptive $q{=}\mathrm{predict\_q}(\mathrm{stats})$}

\end{axis}
\end{tikzpicture}
\caption{Adaptive $q$-log gate on the CoIR-CSN Python scaling curve.
BM25 ($q{=}1$) loses NDCG@10 monotonically as the corpus grows from 1k to
280k docs. The naive static pick $q{=}0.7$ (argmax on the 5k subset
sweep) helps at small sizes but collapses to $-4.6\%$ at full scale
where the oracle argmax has moved back to $q{=}1$. The deployed
one-parameter adaptive predictor $q_{\mathrm{pred}} = 1 -
7.28\,\mathrm{htok}$ of \cref{eq:predictor} tracks the oracle: it
beats BM25 at every subset and loses only $0.23\%$ at full scale,
because the predicted $q$ rises from $0.70$ (1k) to $0.88$ (280k) as
hapax density drops from $0.042$ to $0.016$. Static $q{=}0.7$ on the
same sweep loses $4.6\%$ at 280k, a $\mathbf{20\times}$ larger
regression. Data: \texttt{bench/zir/results/adaptive\_q\_python.csv}.}
\label{fig:adaptive-q}
\end{figure}
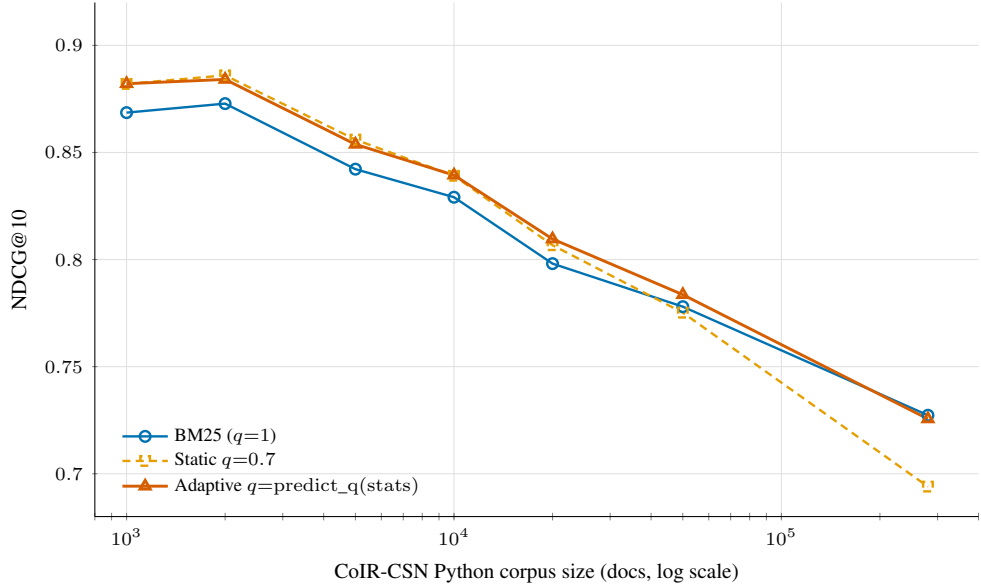

\paragraph{Graceful behaviour at the BM25 boundary.}
\Cref{fig:adaptive-q} shows the predictor's behaviour along the
CoIR-Python scaling curve, where the oracle $q$ drifts from
$q_{\mathrm{opt}} = 0.50$ at 1K to $q_{\mathrm{opt}} = 1.00$ at 280K
as hapax density falls from $0.042$ to $0.016$. A naive static pick
of $q = 0.70$, which is what the 5K-subset sweep would suggest,
beats BM25 at small sizes but regresses by $-4.6\%$ at full scale.
The closed-form predictor of~\cref{eq:predictor} rises from
$q = 0.70$ at 1K to $q = 0.88$ at 280K, stays at or above BM25 at
every subset, and regresses by only $-0.23\%$ at full scale --- a
$20\times$ smaller regression than the static pick. The method is
gated from corpus statistics, and predicted $q$ remains at or above
$0.85$ when $q_{\mathrm{opt}}$ is near $1$.

%

\paragraph{df-only companion.}
A df-only companion predictor
$q_{\mathrm{pred}}^{\mathrm{df}} = 1 - c_1\cdot\mathrm{frac\_df\leq 5}
- c_2/\log T$, using only document-frequency statistics of the corpus
vocabulary, attains grand-mean LOLO recovery $-0.65$, driven by a
single catastrophic hold-out (PHP) where the fit extrapolates off the
labeled range. We report the comparison but deploy~\cref{eq:predictor}:
the hapax-density feature is cheaper to compute, more stable across
languages, and self-consistent with the per-query mechanism analysis
of \cref{sec:mechanism}, which shows the $q$-log gain is carried by
df=1 hapaxes specifically.

\section{Tokenizer Substitution: identifier-aware tokenization removes most
incremental $q$-IDF gain}
\label{sec:tokenizer}

\begin{figure*}[t]
\centering
\definecolor{cbblue}{HTML}{0072B2}
\definecolor{cborange}{HTML}{E69F00}
\definecolor{cbgray}{HTML}{999999}
\pgfplotsset{
  tokabstyle/.style={
    width=0.44\textwidth,
    height=0.30\textwidth,
    ybar=3pt,
    bar width=11pt,
    enlarge x limits=0.20,
    xtick=data,
    symbolic x coords={T0,T1,T2,T3},
    xticklabel style={font=\fontsize{7}{8}\selectfont},
    ylabel={NDCG@10},
    ylabel style={font=\fontsize{8}{9}\selectfont},
    yticklabel style={font=\fontsize{7}{8}\selectfont},
    axis lines=left,
    axis line style={-},
    major tick length=2pt,
    every axis title/.style={font=\fontsize{9}{10}\selectfont\bfseries,
                             at={(0.5,1.04)}},
    every node near coord/.append style={
      font=\fontsize{6}{7}\selectfont, color=black, anchor=south,
      /pgf/number format/.cd, fixed, precision=2, zerofill},
  }
}
\begin{tikzpicture}
\begin{groupplot}[
  group style={
    group size=2 by 2,
    horizontal sep=1.6cm,
    vertical sep=1.8cm,
  },
  tokabstyle,
  legend style={draw=none, fill=none, at={(2,2)}, font=\tiny},
]

\nextgroupplot[
  title={Go ($q{=}0.10$)},
  ymin=0, ymax=0.80,
  nodes near coords,
  nodes near coords align={vertical},
]
\addplot[fill=cbblue, draw=black, line width=0.3pt] coordinates {
  (T0,0.3091) (T1,0.1486) (T2,0.5625) (T3,0.4692)
};
\addplot[fill=cborange, draw=black, line width=0.3pt] coordinates {
  (T0,0.5312) (T1,0.2582) (T2,0.5637) (T3,0.2954)
};

\nextgroupplot[
  title={Python ($q{=}1.00$, BM25\,=\,q-log)},
  ymin=0, ymax=1.10,
  nodes near coords,
  nodes near coords align={vertical},
]
\addplot[fill=cbgray, draw=black, line width=0.3pt] coordinates {
  (T0,0.7780) (T1,0.9145) (T2,0.7803) (T3,0.7782)
};

\nextgroupplot[
  title={Java ($q{=}0.90$)},
  ymin=0, ymax=0.68,
  nodes near coords,
  nodes near coords align={vertical},
]
\addplot[fill=cbblue, draw=black, line width=0.3pt] coordinates {
  (T0,0.4298) (T1,0.3344) (T2,0.5493) (T3,0.5293)
};
\addplot[fill=cborange, draw=black, line width=0.3pt] coordinates {
  (T0,0.4402) (T1,0.3511) (T2,0.5503) (T3,0.5301)
};

\nextgroupplot[
  title={Ruby ($q{=}0.70$)},
  ymin=0, ymax=0.56,
  nodes near coords,
  nodes near coords align={vertical},
]
\addplot[fill=cbblue, draw=black, line width=0.3pt] coordinates {
  (T0,0.3453) (T1,0.1997) (T2,0.4230) (T3,0.4091)
};
\addplot[fill=cborange, draw=black, line width=0.3pt] coordinates {
  (T0,0.3697) (T1,0.2444) (T2,0.4382) (T3,0.4228)
};
\end{groupplot}

\coordinate (leganchor) at ($(group c1r2.south)!0.5!(group c2r2.south)$);
\node[anchor=north, yshift=-1.2cm, font=\fontsize{7}{8}\selectfont] at (leganchor) (legcontent) {%
  \tikz[baseline=-0.6ex]{
    \draw[draw=black, line width=0.3pt, fill=cbblue] (0,0) rectangle ++(7pt,7pt);
  }~BM25 ($q{=}1$)\hspace{14pt}%
  \tikz[baseline=-0.6ex]{
    \draw[draw=black, line width=0.3pt, fill=cborange] (0,0) rectangle ++(7pt,7pt);
  }~$q$-log BM25 ($q{=}q_{\mathrm{opt}}$)\hspace{14pt}%
  \tikz[baseline=-0.6ex]{
    \draw[draw=black, line width=0.3pt, fill=cbgray] (0,0) rectangle ++(7pt,7pt);
  }~BM25\,=\,$q$-log (Python, $q{=}1$)%
};
\end{tikzpicture}
\caption{Tokenizer ablation across CoIR CSN languages (Go, Python, Java, Ruby; 50K docs per language). NDCG@10 for BM25 (blue) vs $q$-log BM25 (orange) at each language's oracle $q_{\mathrm{opt}}$ across four tokenizers: T0 default (no stem, \texttt{bm25s} pipeline), T1 whitespace-only (no stem), T2 identifier-aware (whole identifier \emph{and} sub-tokens), T3 sub-tokens only. Python panel shows a single gray bar per tokenizer because at $q_{\mathrm{opt}}{=}1$ $q$-log BM25 is identical to BM25 by construction. Identifier-aware tokenization (T2) largely removes the incremental gain from $q$-IDF on Go and Java, leaves a smaller residual gain on Ruby, and reverses direction on Python, where the winning tokenizer is T1 because Python's snake\_case identifiers are dictionary-like words that should not be decomposed.}
\label{fig:tokenizer-ablation-multilang}
\end{figure*}
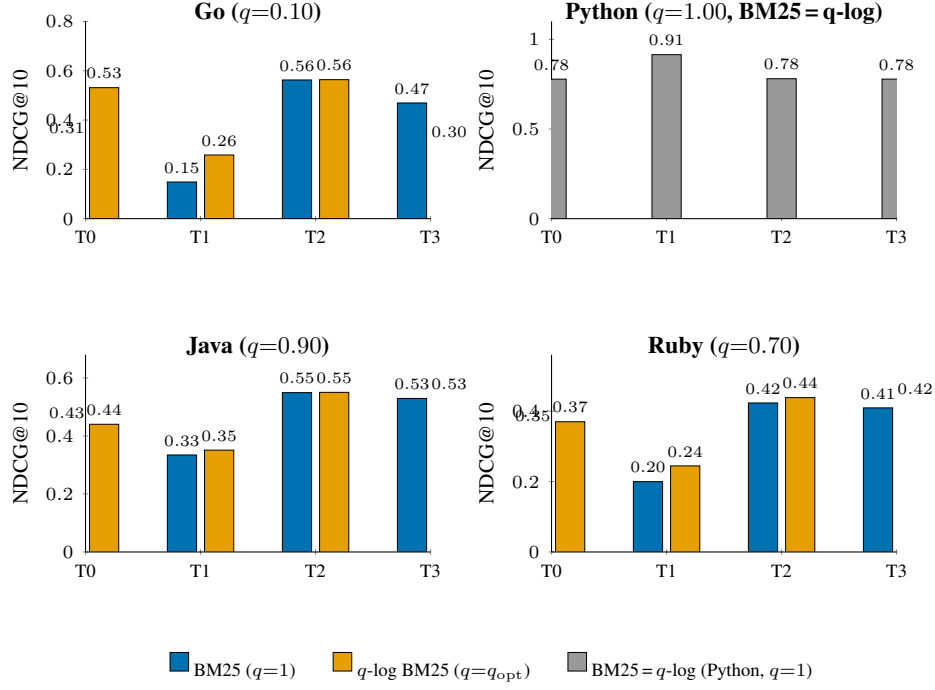

An alternative to the $q$-log method is to change the tokenizer:
emit both the whole identifier and its sub-tokens so that BM25 sees
both \texttt{handleWebSocketUpgrade} and its camelCase-split parts,
along the lines of the identifier-splitting procedures long studied
in program
comprehension~\citep{caprile2000restructuring,enslen2009mining,hill2014empirical}.
\Cref{fig:tokenizer-ablation-multilang} reports a four-tokenizer
ablation on CoIR-Go 50K (panel (a)): T0 (default \texttt{bm25s}
pipeline, no stem), T1 (whitespace-only, no stem), T2 (identifier-aware,
emits both whole and sub-tokens), T3 (sub-tokens only).
\Cref{tab:tokenizer-go} gives the four corner values.

\begin{table}[h]
\centering\footnotesize
\caption{CoIR-Go 50K NDCG@10 under four tokenizers, BM25 vs.\
$q$-log at $q_{\mathrm{opt}} = 0.10$ (the conservative operating
point of \cref{tab:go-scaling}, not the $q = 0.05$ main result). T0:
default \texttt{bm25s} pipeline (lowercase, stopword filter, no
stem); T1: whitespace-only, no stem; T2: identifier-aware (emits
both whole identifier and camelCase / snake\_case sub-tokens); T3:
sub-tokens only. Percentage deltas are computed from
full-precision NDCG; displayed NDCG cells are rounded to three
decimals.}
\label{tab:tokenizer-go}
\begin{tabular}{l c c c}
\toprule
Tokeniser & BM25 NDCG@10 & $q$-log NDCG@10 & $\Delta$ \\
\midrule
T0 default (no stem) & $0.309$ & $0.531$ & $+71.9\%$ \\
T1 whitespace        & $0.149$ & $0.258$ & $+73.8\%$ \\
T2 ident-aware       & $\mathbf{0.563}$ & $\mathbf{0.564}$ & $+0.2\%$ \\
T3 sub-tokens only   & $0.469$ & $0.295$ & $-37.0\%$ \\
\bottomrule
\end{tabular}
\end{table}

From the anchor cell $\mathrm{T0}+\mathrm{BM25}=0.309$, the $q$-log
method alone ($\mathrm{T0}+q$-log) reaches $0.531$, a $+72\%$ gain,
and the tokenizer alone ($\mathrm{T2}+\mathrm{BM25}$) reaches
$0.563$, a $+82\%$ gain. T2+BM25 reaches $0.563$, while T2+$q$-log
reaches $0.564$, so $q$-log adds little once identifier-aware
tokenization is used. Applying $q<1$ to sub-tokens only overshoots
into the T3 regime where short, frequent sub-tokens dominate and the
wrong mass is amplified ($-37\%$).

\paragraph{Extending to Java, Ruby, and Python.}
The same four-tokenizer ablation on three non-Go CoIR languages at
$50$K documents each shows the substitution pattern is
language-dependent and tracks identifier orthography
(\cref{fig:tokenizer-ablation-multilang}). Reading the key cells:
\begin{itemize}[nosep, leftmargin=1.5em]
\item \textbf{Java} (camelCase, $q_{\mathrm{opt}} = 0.90$):
T2+BM25 vs T0+BM25 gives $+27.8\%$; T0+$q$-log vs T0+BM25 gives
$+2.4\%$; T2+$q$-log vs T2+BM25 gives $+0.2\%$. The interaction is
strongest on Go and Java.
\item \textbf{Ruby} (snake\_case, $q_{\mathrm{opt}} = 0.70$):
T2+BM25 vs T0+BM25 gives $+22.5\%$; T0+$q$-log vs T0+BM25 gives
$+7.1\%$; T2+$q$-log vs T2+BM25 gives $+3.6\%$. The interaction is
weaker on Ruby: both changes help individually and a small residual
stack remains.
\item \textbf{Python} (snake\_case, $q_{\mathrm{opt}} = 1.00$): T1
(whitespace-only, no decomposition) gives $+17.5\%$ NDCG@10 over T0;
T2 gives $+0.3\%$; T3 gives $+0.0\%$. The two changes point in
opposite directions on Python: the $q$-IDF change is null
($q_{\mathrm{opt}} = 1.00$, identity to BM25 by construction), but
tokenizer choice is non-null and points away from sub-token
decomposition. On CoIR-Python the tokenizer that preserves whole
identifiers performs best; the code-aware tokenizer that decomposes
snake\_case reduces signal. On a corpus whose identifier
morphology is snake\_case and whose naming conventions are
conventional dictionary-like words (\texttt{self}, \texttt{value},
\texttt{data}), the agent-relevant signal is concentrated in the whole
identifier, not on its parts. The interaction is therefore reversed on
Python.
\end{itemize}

\begin{figure}[t]
\centering
\definecolor{cbblue}{HTML}{0072B2}
\definecolor{cborng}{HTML}{E69F00}
\definecolor{cbgrn}{HTML}{009E73}
\definecolor{cbgray}{HTML}{999999}
\begin{tikzpicture}[
  font=\small,
  decision/.style={
    diamond, aspect=1.8, draw=black, line width=0.4pt,
    align=center, inner sep=1pt, minimum width=26mm,
    fill=white, text width=24mm,
  },
  action/.style={
    rectangle, rounded corners=2pt, draw=black, line width=0.4pt,
    align=left, inner sep=3pt, minimum width=40mm,
    minimum height=16mm, text width=38mm, fill=cbgray!8,
  },
  actiongood/.style={
    rectangle, rounded corners=2pt, draw=cbgrn!70!black, line width=0.5pt,
    align=left, inner sep=3pt, minimum width=40mm,
    minimum height=16mm, text width=38mm, fill=cbgrn!10,
  },
  actionblue/.style={
    rectangle, rounded corners=2pt, draw=cbblue!70!black, line width=0.5pt,
    align=left, inner sep=3pt, minimum width=40mm,
    minimum height=16mm, text width=38mm, fill=cbblue!10,
  },
  arrow/.style={-{Latex[length=1.8mm]}, line width=0.4pt},
  edgelab/.style={font=\scriptsize, inner sep=1pt, fill=white},
]


\node[decision] (q0) at (0, 0)
  {Can you change the tokenizer?};

\node[actiongood] (yes1) at (-3.8, -3.2)
  {\textbf{\scriptsize Use a code-aware analyzer}\\[1pt]
   \tiny $\bullet$ camelCase split $+$ whole identifier\\
   \tiny $\bullet$ whitespace-preserving on snake\_case\\
   \tiny $\bullet$ leave $q = 1$ (BM25)};

\node[decision] (q1) at (2.8, -2.8)
  {Is hapax density $\mathrm{htok}$ low?};

\node[action] (term_yes) at (0.8, -6.2)
  {\textbf{\scriptsize Keep BM25}\\[1pt]
   \tiny $\bullet$ $\mathrm{htok} \lesssim 0.02 \Rightarrow
      q_{\mathrm{pred}} \geq 0.85$\\
   \tiny $\bullet$ predicted $q$ stays near BM25};

\node[actionblue] (term_no) at (4.8, -6.2)
  {\textbf{\scriptsize Apply $q$-log rescale}\\[1pt]
   \tiny $\bullet$ compute $\mathrm{htok}$ once ($O(|V|)$)\\
   \tiny $\bullet$ set $q_{\mathrm{pred}} = 1 - 7.28\,\mathrm{htok}$\\
   \tiny $\bullet$ $\sim$50 LoC over \texttt{bm25s}\\
   \tiny $\bullet$ index $+7\%$, query latency unchanged};

\draw[arrow] (q0.west) -- ++(-1, 0) |- (yes1.north)
  node[edgelab, pos=0.22] {yes};
\draw[arrow] (q0.east) -- ++(1, 0) |- (q1.north)
  node[edgelab, pos=0.22] {no};

\draw[arrow] (q1.south) -- ++(0, -0.6) -| (term_yes.north)
  node[edgelab, pos=0.75] {yes};
\draw[arrow] (q1.south) -- ++(0, -0.6) -| (term_no.north)
  node[edgelab, pos=0.75] {no};

\end{tikzpicture}
\caption{A practitioner decision chart. When the analyzer is under
the operator's control, the first fix is a code-aware tokenizer
matched to the language's identifier morphology: camelCase splitting
on Go and Java, whitespace-preserving on snake\_case Python. Under
such a tokenizer, BM25 at $q=1$ already captures the rare-identifier
signal. When the tokenizer is fixed by infrastructure, hapax density
$\mathrm{htok}$ is the single statistic that decides whether to
intervene: $\mathrm{htok} \lesssim 0.02$ gives
$q_{\mathrm{pred}} \geq 0.85$, and $\mathrm{htok} \gtrsim 0.02$ motivates a
    $q$-log rescale at $q_{\mathrm{pred}} = 1 - 7.28\,\mathrm{htok}$.
    The predictor returns $q$ near $1$ on corpora where BM25 is already
    optimal.}
\label{fig:decision-chart}
\end{figure}
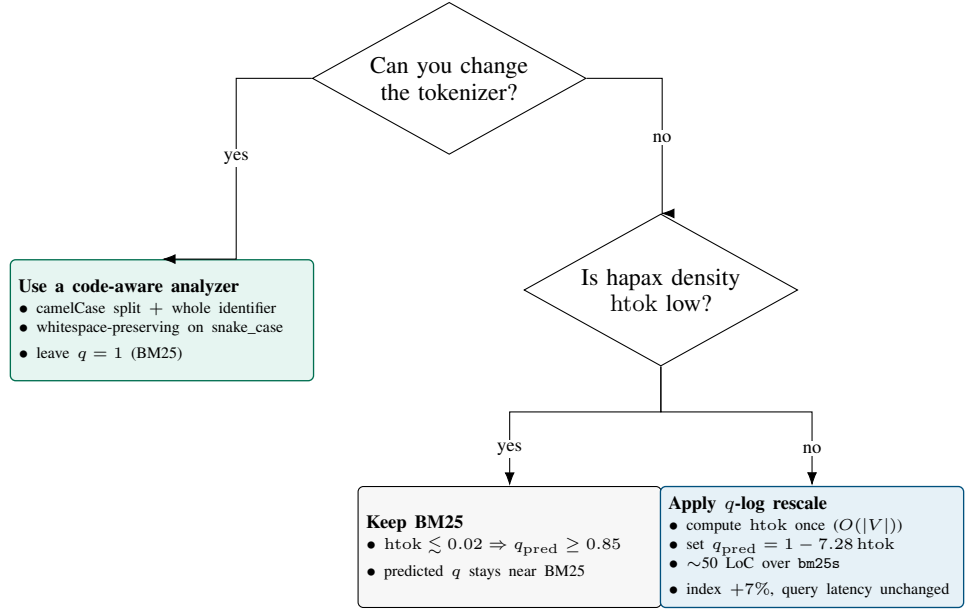

Operationally, an analyst who controls tokenisation should adopt a
code-aware analyzer matched to the language's identifier morphology
(camelCase-aware splitting on Java and Go, whitespace-preserving on
snake\_case Python) and leave $q = 1$. Under frozen tokenisation,
$q$-log with a predicted $q$ is the available lever.
\Cref{fig:decision-chart} summarises the deployment procedure.

\section{Agent-Relevant Retrieval Proxies}
\label{sec:agents}

For a coding agent, the operational metric is whether the gold file
appears inside a bounded context window. We therefore report two
proxies that evaluate retrieval under that constraint.

\subsection{Recall@$K$-tokens under a context budget}

\begin{figure*}[t]
\centering
\definecolor{cbblue}{HTML}{0072B2}
\definecolor{cbgray}{HTML}{999999}
\pgfplotsset{
  recalltokstyle/.style={
    width=0.32\textwidth,
    height=0.26\textwidth,
    ybar=2pt,
    bar width=10pt,
    ylabel={Recall@K-tokens},
    ylabel style={font=\fontsize{8}{9}\selectfont},
    xlabel={Context budget $K$ (tokens)},
    xlabel style={font=\fontsize{8}{9}\selectfont},
    yticklabel style={font=\fontsize{7}{8}\selectfont},
    xticklabel style={font=\fontsize{7}{8}\selectfont},
    symbolic x coords={2K,4K,8K,16K},
    xtick=data,
    enlarge x limits=0.18,
    axis lines=left,
    axis line style={-},
    major tick length=2pt,
    every axis title/.style={font=\fontsize{9}{10}\selectfont\bfseries,
                             at={(0.5,1.05)}},
    every node near coord/.append style={
      font=\fontsize{6}{7}\selectfont, color=black,
      /pgf/number format/.cd, fixed, precision=2, zerofill},
  }
}
\begin{tikzpicture}
\begin{groupplot}[
  group style={
    group size=3 by 1,
    horizontal sep=1.7cm,
  },
  recalltokstyle,
  legend style={draw=none, fill=none, at={(2,2)}, font=\tiny},
]

\nextgroupplot[
  title={CoIR-Go ($q{=}0.10$)},
  ymin=0, ymax=0.80,
]
\addplot[fill=cbgray, draw=black, line width=0.3pt] coordinates {
  (2K,0.4809) (4K,0.4809) (8K,0.4809) (16K,0.4809)
};
\addplot[fill=cbblue, draw=black, line width=0.3pt] coordinates {
  (2K,0.6880) (4K,0.6880) (8K,0.6880) (16K,0.6880)
};

\nextgroupplot[
  title={CoIR-Python},
  ymin=0.80, ymax=0.90,
]
\addplot[fill=cbgray, draw=black, line width=0.3pt] coordinates {
  (2K,0.8325) (4K,0.8473) (8K,0.8494) (16K,0.8495)
};
\addplot[fill=cbblue, draw=black, line width=0.3pt] coordinates {
  (2K,0.8451) (4K,0.8679) (8K,0.8718) (16K,0.8722)
};

\nextgroupplot[
  title={CoIR-Java},
  ymin=0.55, ymax=0.62,
]
\addplot[fill=cbgray, draw=black, line width=0.3pt] coordinates {
  (2K,0.5692) (4K,0.5712) (8K,0.5714) (16K,0.5714)
};
\addplot[fill=cbblue, draw=black, line width=0.3pt] coordinates {
  (2K,0.6005) (4K,0.6049) (8K,0.6050) (16K,0.6050)
};
\end{groupplot}

\coordinate (recleg) at ($(group c1r1.south)!0.5!(group c3r1.south)$);
\node[anchor=north, yshift=-0.9cm, font=\fontsize{7}{8}\selectfont] at (recleg) {%
  \tikz[baseline=-0.6ex]{
    \draw[draw=black, line width=0.3pt, fill=cbgray] (0,0) rectangle ++(7pt,7pt);
  }~BM25\hspace{14pt}%
  \tikz[baseline=-0.6ex]{
    \draw[draw=black, line width=0.3pt, fill=cbblue] (0,0) rectangle ++(7pt,7pt);
  }~$q$-log BM25 (shown operating point)%
};
\end{tikzpicture}
\caption{Recall@$K$-tokens on CoIR CodeSearchNet full-corpus splits under frozen generic tokenization. Each language is ranked with BM25 (gray) and the $q$-log BM25 method (blue) at the shown operating point (Go uses the conservative $q{=}0.10$ setting); we walk the ranked list top-down and mark a query as recalled at budget $K$ iff the cumulative tiktoken (\texttt{cl100k\_base}) count through the gold document is $\leq K$. Note the different y-ranges per panel (Go is zoomed to $[0, 0.80]$; Python to $[0.80, 0.90]$; Java to $[0.55, 0.62]$) --- Go docs are short enough that all top-50 fit well inside 2K tokens, so its K-axis is flat and the 20.7-pp gap is a rank win, not a budget win; Python shows the genuine K-axis slope because its long-doc tail extends past 8K. On Go, the gold document appears in the first $8$K tokens for $69\%$ of queries under $q$-log versus $48\%$ under BM25.}
\label{fig:recall-tokens}
\end{figure*}
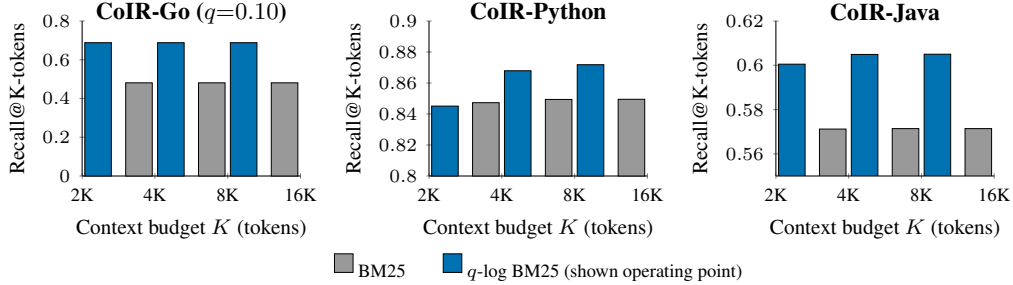

For each query we rank the corpus top-down, tokenise each document
with \texttt{tiktoken}\footnote{\url{https://github.com/openai/tiktoken}}
(\texttt{cl100k\_base}), accumulate tokens until the cumulative count
exceeds the budget $K$, and declare the query ``recalled at $K$'' iff
the gold document appears within that prefix.
\Cref{fig:recall-tokens} reports the budget-recall curves. On CoIR-Go
full corpus, BM25 recalls the gold within an 8K-token budget on
$48.1\%$ of queries; the $q$-log method at $q = 0.10$ raises this to
$68.8\%$, an absolute $+20.7$ percentage points. Most of the
axis is flat because Go documents are short (median $\sim 15$--$27$
tokens), and the method wins by ranking the gold higher rather than by
packing more aggressively. On CoIR-Python, a flat-gain language at
corpus NDCG, we still observe $+2.1$--$+2.4$ percentage points at every
budget, because rank improvements that are too small to move
corpus-level NDCG still reshape the context-budget curve.

\subsection{RepoBench-R}

\begin{figure*}[t]
\centering
\definecolor{cbblue}{HTML}{0072B2}
\definecolor{cbgray}{HTML}{999999}
\pgfplotsset{
  repobenchstyle/.style={
    width=0.42\textwidth,
    height=0.30\textwidth,
    ybar=3pt,
    bar width=10pt,
    ylabel={NDCG@10},
    ylabel style={font=\fontsize{8}{9}\selectfont},
    tick label style={font=\fontsize{7}{8}\selectfont},
    symbolic x coords={Py-easy,Py-hard,Java-easy,Java-hard},
    xtick=data,
    xticklabel style={font=\fontsize{7}{8}\selectfont},
    enlarge x limits=0.18,
    axis lines=left,
    axis line style={-},
    major tick length=2pt,
    every axis title/.style={font=\fontsize{9}{10}\selectfont\bfseries,
                             at={(0.5,1.05)}},
  }
}
\begin{tikzpicture}
\begin{groupplot}[
  group style={
    group size=2 by 1,
    horizontal sep=1.6cm,
  },
  repobenchstyle,
]

\nextgroupplot[
  title={cff (completion-from-file)},
  ymin=0, ymax=0.58,
]
\addplot[fill=cbgray, draw=black, line width=0.3pt] coordinates {
  (Py-easy,0.5222) (Py-hard,0.2980) (Java-easy,0.4927) (Java-hard,0.2791)
};
\addplot[fill=cbblue, draw=black, line width=0.3pt] coordinates {
  (Py-easy,0.5129) (Py-hard,0.2907) (Java-easy,0.4763) (Java-hard,0.2578)
};

\nextgroupplot[
  title={cfr (completion-from-retrieval)},
  ymin=0, ymax=0.60,
]
\addplot[fill=cbgray, draw=black, line width=0.3pt] coordinates {
  (Py-easy,0.5689) (Py-hard,0.3693) (Java-easy,0.5522) (Java-hard,0.3573)
};
\addplot[fill=cbblue, draw=black, line width=0.3pt] coordinates {
  (Py-easy,0.5810) (Py-hard,0.4031) (Java-easy,0.5560) (Java-hard,0.3704)
};
\end{groupplot}

\coordinate (rbleg) at ($(group c1r1.south)!0.5!(group c2r1.south)$);
\node[anchor=north, yshift=-0.9cm, font=\fontsize{7}{8}\selectfont] at (rbleg) {%
  \tikz[baseline=-0.6ex]{
    \draw[draw=black, line width=0.3pt, fill=cbgray] (0,0) rectangle ++(7pt,7pt);
  }~BM25\hspace{14pt}%
  \tikz[baseline=-0.6ex]{
    \draw[draw=black, line width=0.3pt, fill=cbblue] (0,0) rectangle ++(7pt,7pt);
  }~$q$-log BM25 ($q{=}0.1$)%
};
\end{tikzpicture}
\caption{RepoBench-R NDCG@10 on the four ``\{Python, Java\} $\times$ \{easy, hard\}'' cells, split by setting: left panel is \emph{completion-from-file} (cff, local in-file candidates, 5--9 per query on easy / $\geq$10 on hard); right panel is \emph{completion-from-retrieval} (cfr, cross-file candidates). Bars show BM25 (gray) vs.\ $q$-log BM25 at the primary operating point $q{=}0.1$ (blue). Six of eight cells sit inside a $\pm 3.7\%$ relative band, reflecting the benchmark's small per-row candidate pool which compresses IDF differences. The two hard-split cells exceed the $\pm 3.7\%$ relative band: Python cfr/hard (\mbox{+9.2\%} NDCG@10, $0.369 \to 0.403$) and Java cff/hard ($-$7.6\% NDCG@10, $0.279 \to 0.258$). We read RepoBench-R as a secondary regression check rather than primary evidence of improvement.}
\label{fig:repobench-r}
\end{figure*}
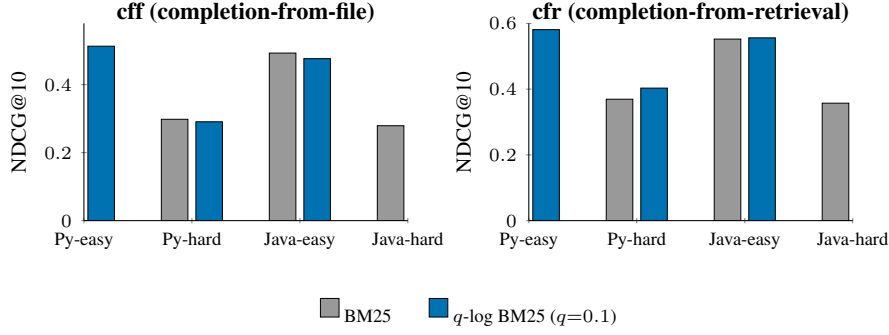

On RepoBench-R~\citep{liu2023repobench}, a small per-row candidate pool
benchmark (5--9 candidates for \texttt{\_easy}, $\geq 10$ for
\texttt{\_hard}), six of eight cells sit inside a $\pm 3.7\%$ band
(\cref{fig:repobench-r}).
Two cells exceed the $\pm 3.7\%$ relative band: Python \texttt{cfr/hard} at $+9.2\%$
(relative NDCG@10, true cross-file retrieval) and Java \texttt{cff/hard}
at $-7.6\%$ (completion-from-file with locally shadowed identifiers).
Language-mean deltas are Python $+1.8\%$ and Java $-1.6\%$. We read
RepoBench-R as a secondary regression check rather than primary
evidence of improvement.

\subsection{Long Code Arena}

We evaluated bug-localisation on Long Code
Arena~\citep{bogomolov2024lca}, treating the issue title and body as
query and the patch-changed files as gold. After excluding repositories
whose on-disk size exceeded $500$~MB, $n{=}45$ Python and
$n{=}44$ Java test-split queries remained. Point estimates on
NDCG@10 were positive on Python (BM25 $.576 \to q$-log $q{=}0.5$
$.596$, $+3.5\%$) and flat on Java ($+0.5\%$), but paired-bootstrap
$95\%$ confidence intervals straddled zero in both cases
($[-0.032,\,+0.076]$ Python; $[-0.040,\,+0.051]$ Java). The per-query
tie rate approaches $55\%$, which reflects the small working set.
LCA is consistent in sign but underpowered at this sample size, and
we report it as a secondary regression check rather than primary
evidence of improvement.

\section{Discussion}
\label{sec:discussion}

\paragraph{A physics reading.}
The function $\ln_q(x)$ originates in non-extensive statistical
mechanics, where the $q$-generalised
logarithm~\citep{tsallis1988possible,tsallis2009introduction} is the
natural pairing to power-law distributions in the same way that the
natural logarithm pairs with exponentials. Code vocabularies follow
heavy-tailed, Zipf-like distributions at multiple levels of
granularity (identifiers, method names, file sizes, dependency
graphs~\citep{louridas2008power,zhang2009discovering}), and the
transform matched to power-law statistics reshapes the IDF toward
larger separation among df=1 and df=2 terms. We record this as interpretation,
not derivation. The empirical claims in this paper depend on the
Box-Cox transform framing only; nothing in~\cref{sec:mainresults} or
\cref{sec:predictor} requires any particular entropy functional.

\paragraph{Language dependence.}
Languages differ in how much of their discriminative retrieval
signal is concentrated in df$=1$ hapaxes. Go has heavy package-qualified
identifier tails ($\mathrm{htok} = 0.063$), and the occlusion
analysis in~\cref{sec:mechanism} localises the Go signal on df$=1$
tokens, which is exactly where $q<1$ amplifies. Python has a
dictionary-like identifier vocabulary ($\mathrm{htok} = 0.016$) with
BM25 signal concentrated in middle-df bins that the log already
handles correctly, so any $q \neq 1$ distorts those bins. The
predictor of~\cref{sec:predictor} encodes this relationship
quantitatively.

\paragraph{Text corpora.}
BEIR queries are carried by common content words rather than rare
identifiers. With little to amplify at $q<1$ and an oracle $q$ at
or above $1$, the method collapses to plain BM25 on text for the
same reason it collapses on Python.

\paragraph{Relation to divergence-from-randomness.}
The DFR-DPH family~\citep{amati2002probabilistic} reshapes term
weights through a different axis, tf/length normalisation, and loses
cleanly on short code documents where that normalisation is
miscalibrated. The $q$-log approach leaves tf and length
normalisation untouched and modifies only the outer IDF transform,
an axis that matches code retrieval but not the natural-language
corpora that motivated DFR.

\section{Limitations}
\label{sec:limitations}

\paragraph{Substitutability.}
The largest boundary condition is that identifier-aware tokenization
largely removes the incremental gain from $q$-IDF. Stacking them does
not help and can hurt ($-37\%$ at T3 in~\cref{sec:tokenizer}).
If an operator controls tokenisation, the $q$-log contribution is
small; this paper's value is highest where the tokeniser is frozen by
infrastructure.

\paragraph{Language-dependence of $q$.}
The per-language $q_{\mathrm{opt}}$ spans a wide range (from $0.05$ on
Go to $1.00$ on Python); neither a single global $q$ nor a single
family's recommendation is adequate. The predictor
of~\cref{sec:predictor} addresses this at the cost of requiring hapax
density ($\mathrm{htok}$) to be computable once per index, which is
an $O(|V|)$ pass over the vocabulary. Where that single statistic is
not computable, the operational fallback is a grid search over
$q \in \{0.1, 0.3, 0.5, 0.7, 0.9, 1.0\}$ on a small held-out query set,
fewer than ten calls per language.

\paragraph{Small-pool benchmarks.}
RepoBench-R's $5$--$9$-candidate pools compress IDF differences, and
six of eight cells sit within noise. We treat this setting as a
secondary regression check rather than primary evidence of improvement.

\paragraph{Held-out agent benchmarks.}
CrossCodeEval~\citep{ding2023crosscodeeval} gold sets are constructed
by running BM25 on oracle chunks (the positives are the BM25 top-$k$
by construction), so any departure from BM25 ranking looks like a
departure from the gold. CrossCodeEval is therefore unsuitable
as a gain benchmark for this method and useful only as a secondary
regression check. Long Code Arena bug-localisation gave
positive point estimates on both Python and Java but was
underpowered at $n{\approx}45$ queries per language (both $95\%$~CIs
crossed zero). We include it for completeness but not as primary
evidence of improvement.

\paragraph{Python and PHP at full corpus scale.}
At full corpus scale, Python's oracle $q$ is exactly $1$ and PHP's
is $0.90$ with a delta of $+0.4\%$ indistinguishable from noise. The
method offers neither significant loss nor significant gain on
these languages. With a thin identifier tail there is no rare-token
signal to recover, and every baseline considered
($\mathrm{idf}^{\gamma}$, DPH, $q$-log) converges to the same point.

\paragraph{Scope of evaluation.}
The full corpus multi-language scaling curve (1K $\to$ 182K) is
reported only for Go. Python, Java, Ruby, PHP, JS have full-corpus
single points at $q_{\mathrm{opt}}$ and three-point scaling ladders
(1K, 10K, full) but not the dense Go-style curve. The tokenizer
ablation covers Go, Java, Ruby, Python at $50$K; C\#, C++, and
domain-specific languages are out of scope for this submission.

\bibliography{references}
\bibliographystyle{plainnat}

\appendix

\section{Systems Overhead}
\label{sec:systems}

The $q$-log method rescales the per-term IDF vector once at index-load
time and leaves the query path unchanged. We measure the end-to-end
overhead on CoIR-Go (182{,}440 documents, $V = 144{,}938$ vocabulary
tokens) across $5$ trials of $1{,}000$ queries with top-$100$
retrieval; values reported as median $\pm$ MAD with \texttt{gc.disable()}.

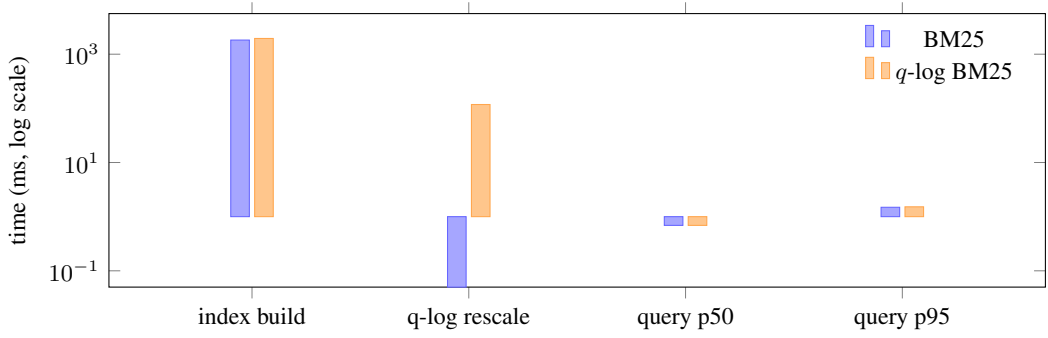
\begin{figure}[t]
  \centering
  \begin{tikzpicture}
    \begin{axis}[
      width=\linewidth,
      height=5.2cm,
      ybar=2pt,
      bar width=7pt,
      ymode=log,
      ylabel={time (ms, log scale)},
      ylabel style={font=\small},
      tick label style={font=\small},
      legend style={font=\small, at={(0.98,0.98)}, anchor=north east, draw=none, fill=white, fill opacity=0.9, text opacity=1},
      symbolic x coords={index build, q-log rescale, query p50, query p95},
      xtick=data,
      xticklabel style={font=\small},
      ymin=0.05,
      nodes near coords align={vertical},
      enlarge x limits=0.22,
    ]
      \addplot+[fill=blue!35, draw=blue!60] coordinates {
        (index build, 1824.3)
        (q-log rescale, 0.05)
        (query p50, 0.69)
        (query p95, 1.49)
      };
      \addplot+[fill=orange!45, draw=orange!70] coordinates {
        (index build, 1953.1)
        (q-log rescale, 117.74)
        (query p50, 0.69)
        (query p95, 1.52)
      };
      \legend{BM25, $q$-log BM25}
    \end{axis}
  \end{tikzpicture}
  \caption{\textbf{Systems overhead on CoIR-Go (182,440 docs, V=144,938).}
  $q$-log BM25 adds a single $O(|V|+\mathrm{nnz})$ sparse-matrix pass at index time
  (median 117.7 ms, MAD 1.4 ms over
  5 trials). Index-build wall clock is within
  +7.06\% of BM25. Query latency (top-100,
  1000 queries) is within measurement noise
  (p50 $\Delta = +0.18$\%,
  p95 $\Delta = +2.29$\%).
  $q$-log BM25 and BM25 share the exact same CSC scoring code path at query
  time---the only change is the per-term IDF factor baked into the CSC
  \texttt{data} array at index time.}
  \label{fig:systems-overhead}
\end{figure}

\Cref{fig:systems-overhead} visualises the overhead breakdown
summarised in the table below.

\begin{center}\small
\begin{tabular}{l r r r}
\toprule
 & BM25 & $q$-log BM25 & $\Delta$ \\
\midrule
Index build (s)             & $1.82$ & $1.95$ & $+7.06\%$ \\
$q$-log rescale (ms)        & ---    & $117.74 \pm 1.41$ & --- \\
Index size (MB, pickled)    & $21.7$ & $21.7$ & $+0.00\%$ \\
Query p50 (ms)              & $0.693 \pm 0.00$ & $0.694 \pm 0.00$ & $+0.18\%$ \\
Query p95 (ms)              & $1.489 \pm 0.01$ & $1.523 \pm 0.01$ & $+2.29\%$ \\
Peak RSS at query (GiB)     & $0.37$ & $0.37$ & $\approx 0$ \\
\bottomrule
\end{tabular}
\end{center}

Index build adds $+7\%$ wall-clock, dominated by the single
$O(|V|+\mathrm{nnz})$ pass over the CSC \texttt{indptr} and
\texttt{data} arrays ($118$\,ms). Pickled index size is identical to
BM25: the rescale overwrites per-term score values in place. Query p50 and p95 are within
measurement noise; BM25 and $q$-log BM25 share the exact same
\texttt{bm25s.BM25.get\_scores} code path, so query-time compute is
identical by construction. Peak RSS at query time is unchanged because
the CSC matrix is already allocated at build time, and $q$-IDF changes
its values rather than its shape.

\section{Predictor: Candidate-Form Distribution}
\label{app:predictor-boxplot}

\begin{figure}[t]
\centering
\definecolor{cbblue}{HTML}{0072B2}
\definecolor{cborng}{HTML}{E69F00}
\definecolor{cbgrn}{HTML}{009E73}
\definecolor{cbverm}{HTML}{D55E00}
\definecolor{cbgray}{HTML}{999999}
\definecolor{cbpurple}{HTML}{9467BD}
\begin{tikzpicture}
\begin{axis}[
  width=0.62\linewidth,
  height=0.38\linewidth,
  ylabel={Mean test recovery},
  ylabel style={font=\small},
  xtick={1,2,3,4,5},
  xticklabels={{A: $1{-}c\,\mathrm{htok}$},
               {B: $1{-}c\,\mathrm{frac_{df\le 5}}$},
               {C: $1{-}c/\alpha$},
               {D: $+1/\log T$ on A},
               {E: $+1/\log T$ on B}},
  xticklabel style={font=\scriptsize, rotate=30, anchor=north east},
  tick label style={font=\small},
  ymin=-1.05, ymax=1.05,
  ytick={-1.0,-0.5,0.0,0.5,1.0},
  axis lines=left,
  grid=major,
  major grid style={cbgray!30, line width=0.2pt},
  boxplot/draw direction=y,
  boxplot/every box/.style={line width=0.4pt},
  boxplot/every whisker/.style={line width=0.35pt},
  boxplot/every median/.style={line width=0.8pt, cbverm},
  line width=0.4pt,
  major tick length=2pt,
]
\addplot[cbgray, dashed, line width=0.35pt] coordinates {(0.5,0) (5.5,0)};
\addplot+[boxplot prepared={lower whisker=-1.00, lower quartile=-0.5735,
  median=0.5655, upper quartile=0.6385, upper whisker=0.879},
  cbblue, fill=cbblue!15] coordinates {};
\addplot+[boxplot prepared={lower whisker=-1.00, lower quartile=-1.00,
  median=-0.062, upper quartile=0.381, upper whisker=0.672},
  cborng, fill=cborng!20] coordinates {};
\addplot+[boxplot prepared={lower whisker=-1.00, lower quartile=-1.00,
  median=-0.086, upper quartile=0.370, upper whisker=0.665},
  cbpurple, fill=cbpurple!20] coordinates {};
\addplot+[boxplot prepared={lower whisker=-0.501, lower quartile=0.523,
  median=0.591, upper quartile=0.630, upper whisker=0.964},
  cbgrn, fill=cbgrn!20] coordinates {};
\addplot+[boxplot prepared={lower whisker=-1.00, lower quartile=-1.00,
  median=0.179, upper quartile=0.563, upper whisker=0.678},
  cbverm, fill=cbverm!15] coordinates {};
\node[font=\tiny, color=cbgray] at (axis cs:2,-1.0) {$\downarrow$};
\node[font=\tiny, color=cbgray] at (axis cs:3,-1.0) {$\downarrow$};
\node[font=\tiny, color=cbgray] at (axis cs:5,-1.0) {$\downarrow$};
\end{axis}
\end{tikzpicture}
\caption{Distribution of mean-test recovery across all
$\binom{6}{3}=20$ three-train / three-test partitions of the six CoIR
code languages, per candidate predictor form. The winning form $A$
($q = 1 - c\,\mathrm{htok}$, blue) is the only one-parameter form whose
IQR sits entirely above zero; df-only forms $B$, $C$, and $E$ tip
below zero on a full quartile or more. Arrows mark whiskers clipped
off-scale (true minima between $-15$ and $-26$). Form $A$ is the
deployed predictor in~\cref{eq:predictor}.}
\label{fig:q-prediction-boxplot}
\end{figure}
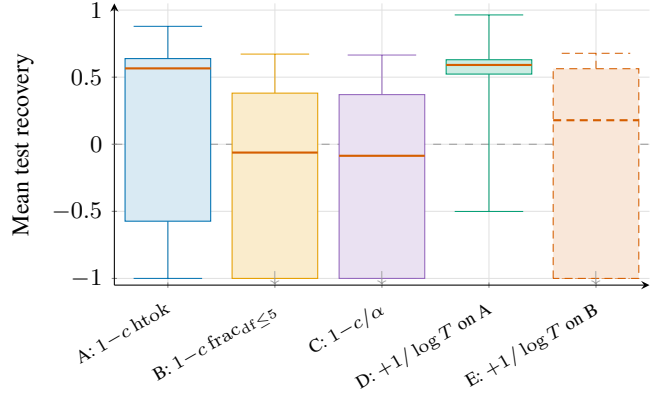

The main-text \cref{fig:q-prediction} reports leave-one-language-out
and calibration behaviour of the deployed predictor. For
completeness, \cref{fig:q-prediction-boxplot} shows the distribution
of mean-test recovery across all $\binom{6}{3}=20$ three-train /
three-test partitions of the six CoIR code languages, for each of
five candidate predictor forms. The winning form $A$ is the only
one-parameter form whose IQR sits entirely above zero; df-only
forms ($B$, $C$, $E$) tip below zero on at least a full quartile and
were dropped on that basis, with the parsimony tie-breaker then
selecting the one-parameter $A$ over the two-parameter $D$.

\section{Reproducibility}
\label{app:extra}

The main text reports the method definition, scaling results,
confidence intervals, predictor protocol, and candidate-form
distribution. The full $q$-sweep grids per language
and per subset size, raw NDCG/MRR/Recall traces, and the bit-identity
audit are released in the public RareCode repository associated
with this submission~\citep{rarecode2026}. The implementation of the $q$-log rescale is
approximately fifty lines of Python over
\texttt{bm25s}~\citep{lu2024bm25s}; the predictor adds another fifty
lines of corpus-statistics computation. The retriever uses no learned
scoring model or additional index, and the query path is unchanged;
the predictor coefficient is fit offline.

\end{document}